\documentclass[twocolumn,showpacs,preprintnumbers,amsmath,amssymb,floatfix]{revtex4}
\usepackage{bm}		% bold math
\usepackage{graphicx}	% figures
\usepackage{dcolumn}	% align columns on decimal point in tables
\begin{document}

\title[Inside charged black holes]{Inside charged black holes II. Baryons plus dark matter}

\author{Andrew J S Hamilton}
\email{Andrew.Hamilton@colorado.edu}
\homepage{http://casa.colorado.edu/~ajsh/}
\author{Scott E Pollack}
\email{pollacks@jilau1.colorado.edu}
\homepage{http://onehertz.colorado.edu/}
\affiliation{
JILA and Dept.\ Astrophysical \& Planetary Sciences,
Box 440, U. Colorado, Boulder CO 80309, USA
}

\newcommand{\dd}{d}
\newcommand{\gvec}{\bm{g}}
\newcommand{\pvec}{\bm{p}}
\newcommand{\gammavec}{\bm{\gamma}}
\newcommand{\xivec}{\bm{\xi}}
\newcommand{\ucoord}{\upsilon}
\newcommand{\utet}{u}
\newcommand{\partialvec}{\bm{\partial}}
\newcommand{\lnt}{\ln\!t}
\newcommand{\perpperp}{\perp\!\!\perp}
\newcommand{\Mc}{M_c}
\newcommand{\vd}{\textsl{v}_d}

\newcommand{\rnorm}{r}
\newcommand{\alphanorm}{\alpha}
\newcommand{\betanorm}{\beta}
\newcommand{\gammanorm}{\gamma}

\newcommand{\grarb}{\gprime}
\newcommand{\hrarb}{\hprime}
\newcommand{\rrarb}{\rprime}
\newcommand{\alphararb}{\alphaprime}
\newcommand{\betararb}{\betaprime}
\newcommand{\gammararb}{\gammaprime}
\newcommand{\lambdararb}{\lambdaprime}
\newcommand{\partialrarb}{\partialprime}

\newcommand{\rgen}{\rprimeprime}
\newcommand{\alphagen}{\alphaprimeprime}
\newcommand{\betagen}{\betaprimeprime}
\newcommand{\gammagen}{\gammaprimeprime}
\newcommand{\lambdagen}{\lambdaprimeprime}
\newcommand{\mugen}{\muprimeprime}
\newcommand{\nugen}{\nuprimeprime}
\newcommand{\partialgen}{\partialprimeprime}

\newcommand{\gtilde}{{\widetilde{g}}}
\newcommand{\htilde}{{\widetilde{h}}}
\newcommand{\rtilde}{{\widetilde{r}}}
\newcommand{\alphatilde}{\widetilde{\alpha}}
\newcommand{\betatilde}{\widetilde{\beta}}
\newcommand{\gammatilde}{\widetilde{\gamma}}
\newcommand{\lambdatilde}{\widetilde{\lambda}}

\newcommand{\gprime}{{g^\prime}}
\newcommand{\hprime}{{h^\prime}}
\newcommand{\rprime}{{r^\prime}}
\newcommand{\alphaprime}{{\alpha^\prime}}
\newcommand{\betaprime}{{\beta^\prime}}
\newcommand{\gammaprime}{{\gamma^\prime}}
\newcommand{\lambdaprime}{{\lambda^\prime}}
\newcommand{\muprime}{{\mu^\prime}}
\newcommand{\nuprime}{{\nu^\prime}}
\newcommand{\partialprime}{\partial^\prime}

\newcommand{\rprimeprime}{{r^{\prime\prime}}}
\newcommand{\alphaprimeprime}{{\alpha^{\prime\prime}}}
\newcommand{\betaprimeprime}{{\beta^{\prime\prime}}}
\newcommand{\gammaprimeprime}{{\gamma^{\prime\prime}}}
\newcommand{\lambdaprimeprime}{{\lambda^{\prime\prime}}}
\newcommand{\muprimeprime}{{\mu^{\prime\prime}}}
\newcommand{\nuprimeprime}{{\nu^{\prime\prime}}}
\newcommand{\partialprimeprime}{\partial^{\prime\prime}}

\hyphenpenalty=3000

%--------------------
% FIG
\newcommand{\varsDMfig}{
    \begin{figure}
    \includegraphics[scale=.6]{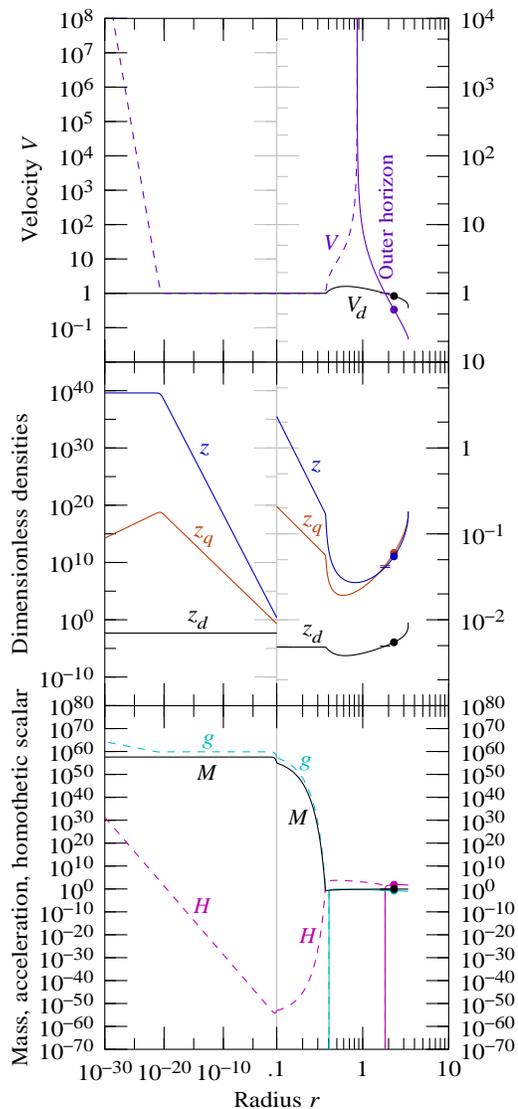}
    \caption[1]{
    \label{varsDM}
(Color online)
A black hole with the same parameters,
an accretion rate $\eta_s = 0.1$ and a charge-to-mass $Q / \Mc = 0.8$,
as that shown in Figure~4 of Paper~1,
except that here the black hole accretes dark matter in addition to baryons,
with density ratio $\rho_d / \rho_b = 0.1$ at the outer sonic point.
Quantities are plotted against radius $r$
in units where the charge-augmented mass at the outer sonic point is unity,
$\Mc = 1$.
To reveal more detail,
the radial axis is split into two regimes with different
horizontal and vertical scales.
Lines are dashed where quantities are negative.
Disks mark the outer sonic point,
where $V = \sqrt{w} \approx 0.577$,
at which the boundary conditions are set.
Short horizontal bars mark the horizon, where $V = 1$.
(Upper panel)
Proper velocity $V$ of the similarity frame
relative to the baryonic frame.
(Middle panel)
Dimensionless proper baryonic mass and charge densities
$z \equiv 4 \pi r^2 \rho_b$
and
$z_q \equiv 4 \pi r^2 q$,
and dimensionless proper dark matter mass density
$z_d \equiv 4 \pi r^2 \rho_d$.
(Bottom panel)
Interior mass $M$,
proper acceleration $g$ experienced by the baryonic fluid,
and the homothetic scalar $H$.
The exponential increase of mass $M$
over a modest range of radii above the inner horizon,
$r \sim 0.4$--$0.1$,
is the signature of mass inflation.
    }
    \end{figure}
}

%--------------------
% FIG
\newcommand{\mrfig}{
    \begin{figure}
    \includegraphics[scale=.7]{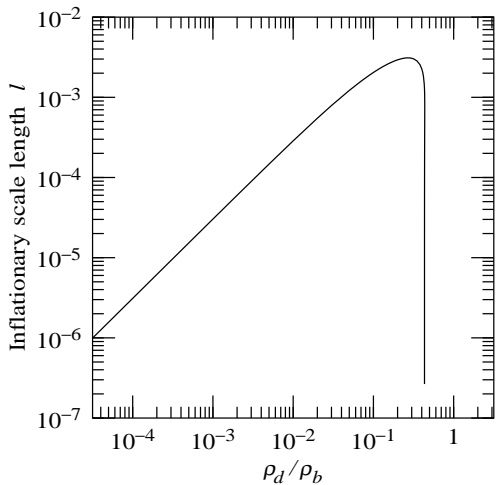}
    \caption[1]{
    \label{mr}
Dimensionless exponential inflationary scale length $l$
as a function of the ratio
$\rho_d / \rho_b$
of dark matter to baryonic proper mass densities at the outer sonic point,
for black holes accreting at rate $\eta_s = 0.1$
and with charge-to-mass $Q / \Mc = 0.8$.
Shorter scale lengths $l$ signify more extreme mass inflation.
    }
    \end{figure}
}

%--------------------
% FIG
\newcommand{\varsWMfig}{
    \begin{figure}
    \includegraphics[scale=.6]{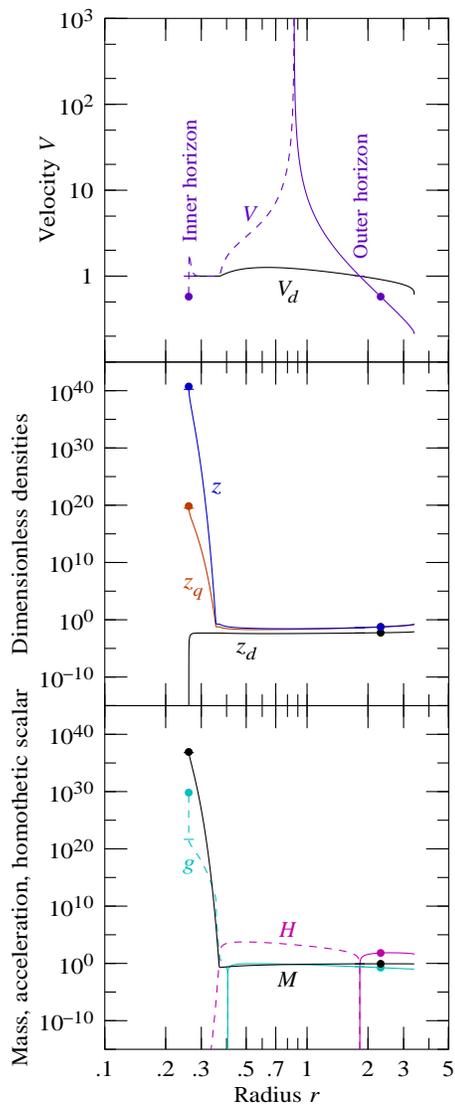}
    \caption[1]{
    \label{varsWM}
(Color online)
Similar to Figure~\protect\ref{varsDM},
but for a black hole in which the absorption rate
of dark matter by baryons is effectively infinite
above a certain large collision energy.
Disks mark sonic points, where $V = \pm \sqrt{w}$.
Short horizontal bars mark horizons, where $V = \pm 1$.
Mass inflation begins,
but is cut short as soon as all the dark matter is absorbed.
With the dark matter gone,
the baryons almost immediately drop through the Cauchy horizon.
The similarity solution terminates inside the Cauchy horizon
at an irregular sonic point.
    }
    \end{figure}
}

%%--------------------
%% FIG
%\newcommand{\shkWMfig}{
%    \begin{figure}
%    \includegraphics[scale=.6]{shk_WM.ps}
%    \caption[1]{
%    \label{shkWM}
%(Color online)
%(Top panel)
%Putative continuations of the
%velocity $V$ inside the Cauchy horizon
%for the model shown in the right panel of Figure~\protect\ref{varsDM},
%which has $Q/\Mc = 0.8$ and zero conductivity.
%The Cauchy horizon, where $V = -1$, is marked by a short horizontal bar.
%A shock decelerates the baryonic fluid from supersonic to subsonic,
%through the sound speed where $V = - \sqrt{w} \approx 0.577$.
%Three cases are shown,
%ranging the gamut
%from an extremely weak shock to an extremely strong shock.
%Dotted lines connect the pre-shock and post-shock velocities.
%In all cases
%the shocked fluid continues downstream for a short distance
%before terminating at an irregular sonic point, marked by a disk,
%where the proper acceleration diverges
%and the similarity solution cannot be continued.
%(Bottom panel)
%Proper acceleration $y \equiv g r$ experienced by the baryonic fluid
%in the vicinity of each of the three putative shocks.
%The fluid decelerates
%(negative, dashed lines)
%pre-shock,
%and accelerates
%(positive, continuous lines)
%post-shock.
%Note that outward = inward!
%In all cases the inward acceleration of the post-shock fluid
%increases without bound, diverging at an irregular sonic point
%where the velocity attempts to accelerate back above the speed of sound.
%    }
%    \end{figure}
%}

%--------------------
% FIG
\newcommand{\appearfig}{
    \begin{figure*}
    \includegraphics[scale=.525]{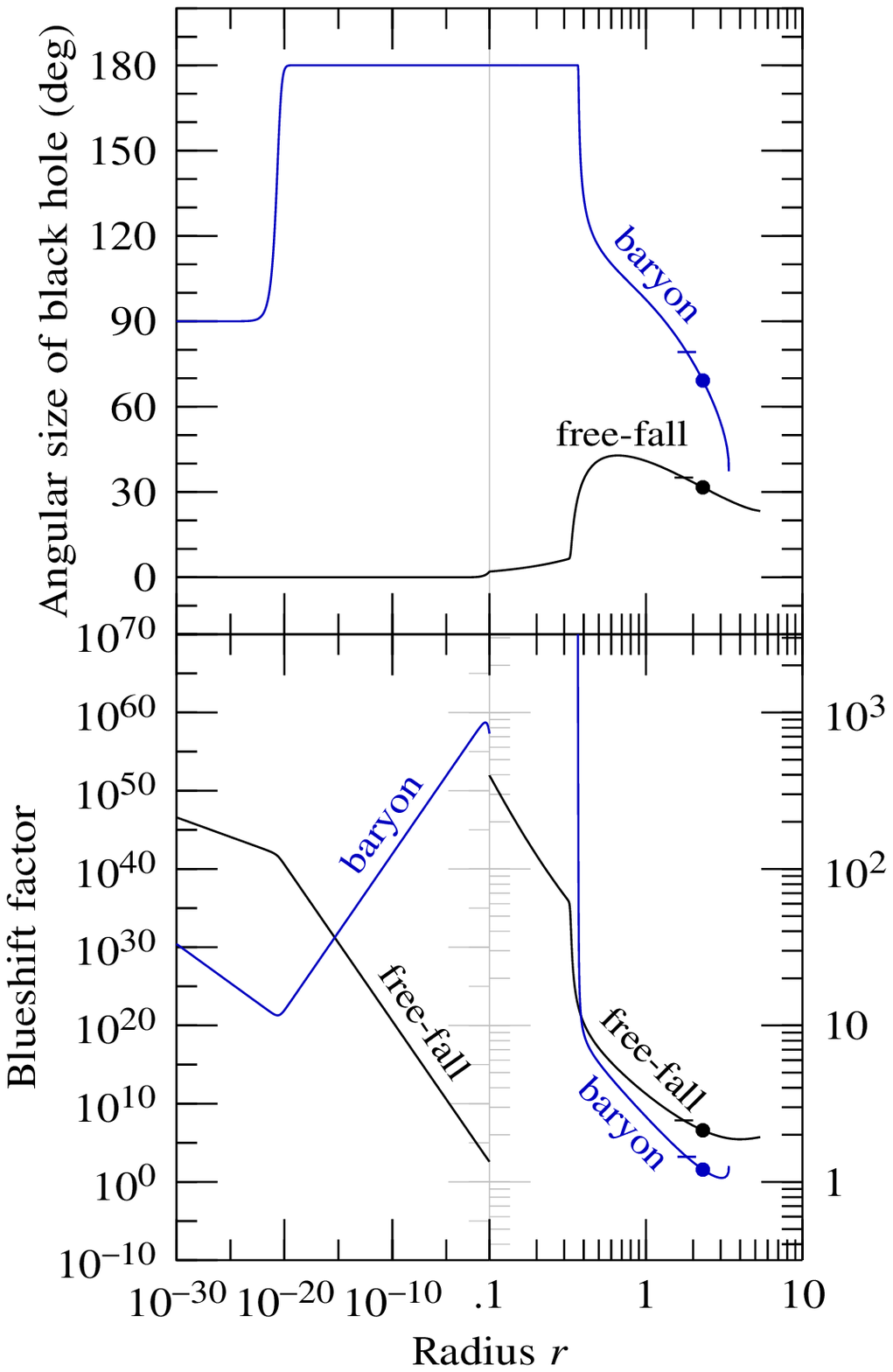}
    \includegraphics[scale=.525]{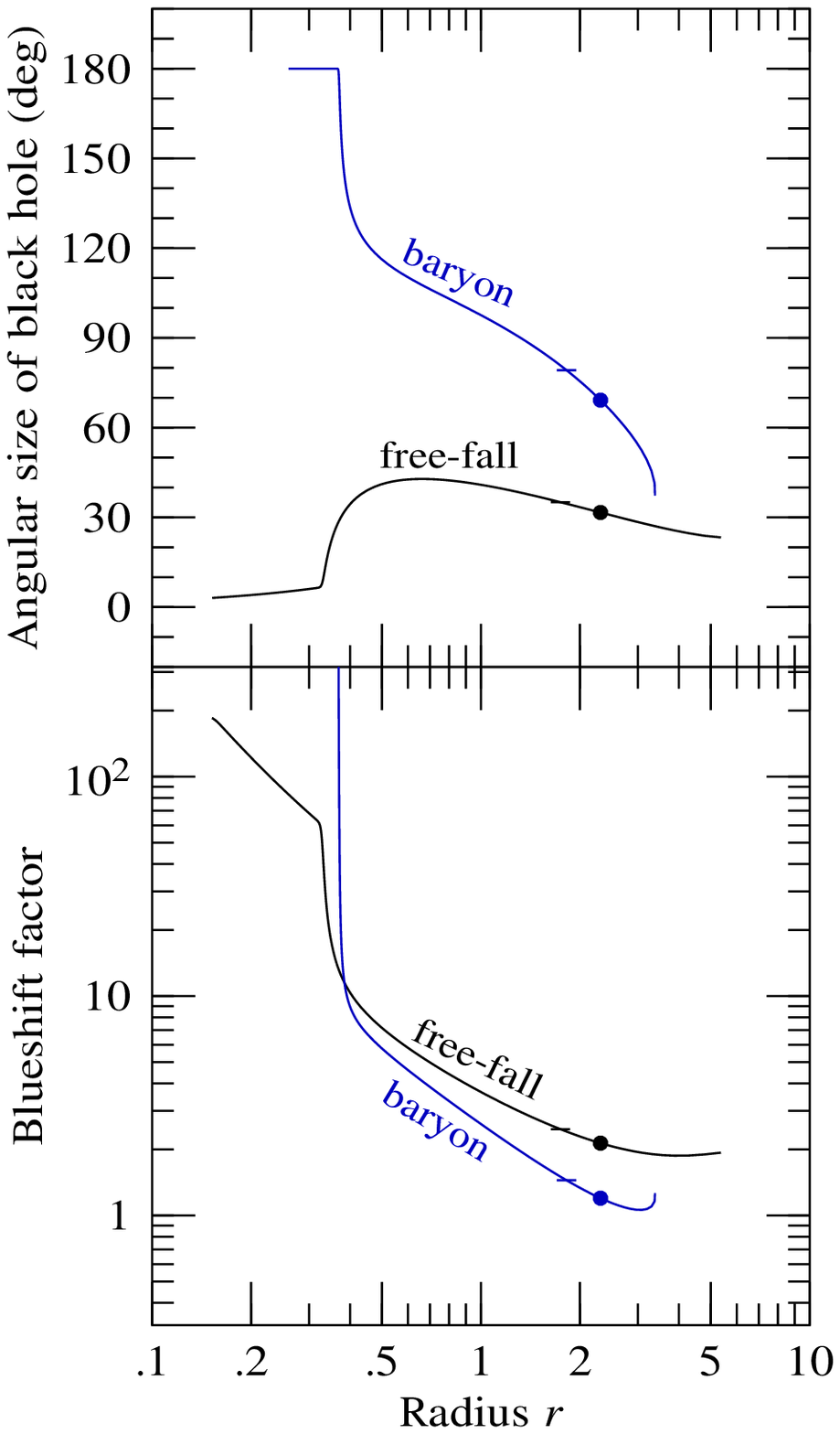}
    \caption[1]{
    \label{appear}
(Color online)
Angular size $\chi_\textrm{ph}$ of the black hole
and the blueshift of photons at the edge of the black hole
perceived by observers in either the baryonic frame
or the free-fall dark matter frame, for
(left) the model
of Figure~\protect\ref{varsDM},
where the black hole accretes non-interacting dark matter,
and (right) the model
of Figure~\protect\ref{varsWM},
where the black hole accretes dark matter
whose cross-section for absorption by baryons is effectively infinite
at high energy.
Light from the outside universe is visible only from outside
the Cauchy horizon,
so lines terminate infinitesimally outside the Cauchy horizon
even in the model at right, in which the baryons drop inside the Cauchy horizon.
    }
    \end{figure*}
}

%--------------------
% FIG
\newcommand{\radDMfig}{
    \begin{figure}
    \includegraphics[scale=.6]{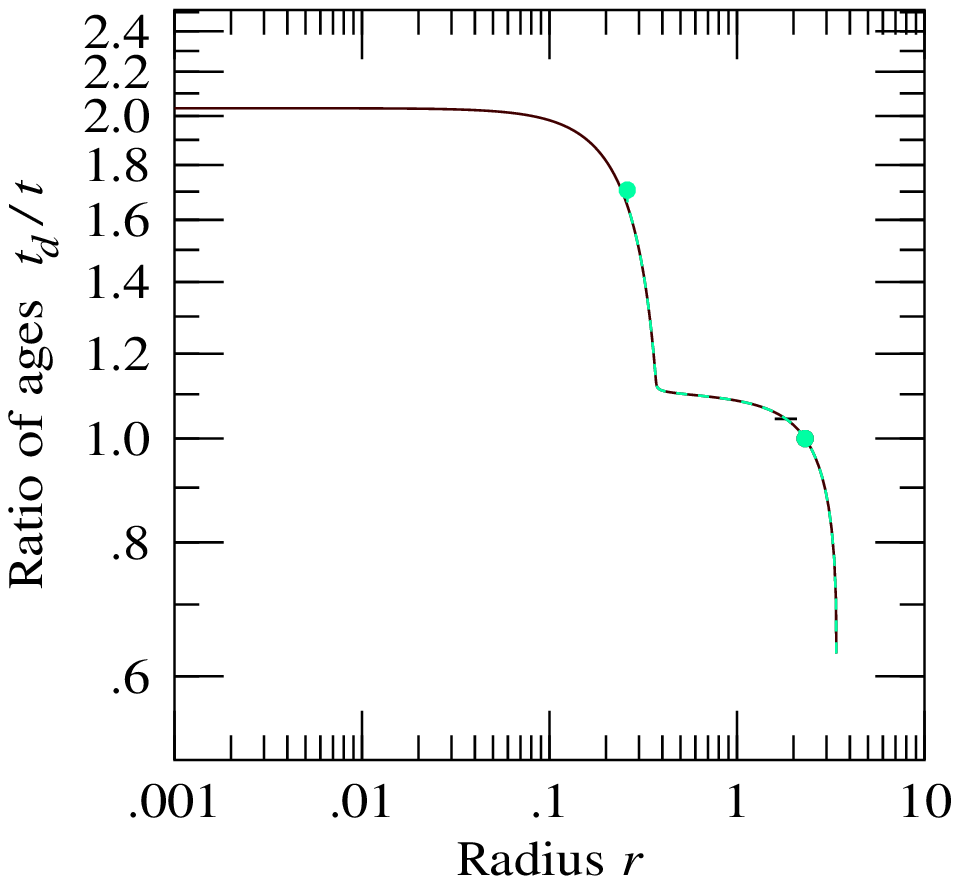}
    \caption[1]{
    \label{radDM}
(Color online)
Ingoing dark matter and outgoing baryons
streaming through each other at radius $r$ inside the black hole
were accreted by the black hole at different times.
The graph shows the ratio $t_d/t$ of ages of the black hole
when the dark matter versus baryons were accreted.
For example, a ratio $t_d/t = 2$ means that
the black hole was twice as old when it accreted the dark matter
as it was when it accreted the baryons.
The solid line is for the model with non-interacting dark matter
shown in Figure~\protect\ref{varsDM},
while the dashed line almost coincident with the solid line
is for the model shown in Figure~\protect\ref{varsDM},
in which the dark matter has an effectively infinite cross-section
for absorption by baryons at high energy.
    }
    \end{figure}
}

\begin{abstract}
This is the second of two companion papers on the interior structure of
self-similar accreting charged black holes.  In the first paper, the black hole
was allowed to accrete only a single fluid of charged baryons.  In this second
paper, the black hole is allowed to accrete in addition a neutral fluid of
almost non-interacting dark matter.  Relativistic streaming between outgoing
baryons and ingoing dark matter leads to mass inflation near the inner horizon.
When enough dark matter has been accreted that the center of mass frame near
the inner horizon is ingoing, then mass inflation ceases and the fluid
collapses to a central singularity.  A null singularity does not form on the
Cauchy horizon.  Although the simultaneous presence of ingoing and outgoing
fluids near the inner horizon is essential to mass inflation, reducing one or
other of the ingoing dark matter or outgoing baryonic streams to a trace
relative to the other stream makes mass inflation more extreme, not the other
way round as one might naively have expected.  Consequently, if the dark matter
has a finite cross-section for being absorbed into the baryonic fluid, then the
reduction of the amount of ingoing dark matter merely makes inflation more
extreme, the interior mass exponentiating more rapidly and to a larger value
before mass inflation ceases.  However, if the dark matter absorption
cross-section is effectively infinite at high collision energy, so that the
ingoing dark matter stream disappears completely, then the outgoing baryonic
fluid can drop through the Cauchy horizon.  In all cases, as the baryons and
the dark matter voyage to their diverse fates inside the black hole, they only
ever see a finite amount of time pass by in the outside universe.  Thus the
solutions do not depend on what happens in the infinite past or future.  We
discuss in some detail the physical mechanism that drives mass inflation.
Although the gravitational force is inward, inward means opposite direction
for ingoing and outgoing fluids near the inner horizon.  Mass inflation is
driven by a feedback loop in which the general relativistic contribution to
the gravitational force sourced by the radial pressure accelerates the ingoing
and outgoing fluids through each other, which increases the radial pressure,
which increases the gravitational force.
%What drives inflation is the general relativistic contribution to
%the gravitational force sourced by the radial pressure.
%The gravitational force accelerates the ingoing and outgoing fluids
%through each other, which increases the radial pressure,
%which increases the gravitational force.
\end{abstract}

	% Classical general relativity

\date{\today}

\maketitle

\section{Introduction}

This is the second of two companions papers
on similarity solutions for the interior structure of
spherically symmetric charged black holes.
%In the first paper, hereafter Paper~1 \cite{Paper1},
%the black hole was allowed to accrete
%a relativistic ($p_b / \rho_b = 1/3$) fluid of charged baryons.
%In the present paper,
%the black hole is allowed in addition to accrete
%a pressureless fluid of non-interacting, or almost non-interacting,
%neutral dark matter.
%The principal aim is to investigate the phenomenon of mass inflation,
%first proposed in a seminal paper by Poisson \& Israel (1990) \cite{PI90}.

In a seminal paper,
Poisson \& Israel (1990)
\cite{PI90}
showed that if ingoing and outgoing fluids are allowed to pass through
each other inside a charged black hole,
then the generic consequence is
`mass inflation'
as the counter-streaming fluids approach the inner horizon.
During mass inflation,
the interior mass, the Misner-Sharp mass \cite{MS64},
a gauge-invariant scalar quantity,
exponentiates to an enormous value.
The phenomenon of mass inflation has been confirmed
analytically and numerically
in many papers
\cite{Ori91,BDIM94,BS95,Burko97,BO98,Burko03,Dafermos04}.

In the first paper of this pair, hereafter Paper~1 \cite{Paper1},
the black hole was allowed to accrete
a relativistic fluid of charged baryons.
The baryons did not undergo mass inflation,
precisely because, by construction,
the single component baryonic fluid considered there
was either ingoing or outgoing, not both.

In the present paper
we allow the black hole to accrete,
in addition to a charged baryonic fluid,
a pressureless neutral `dark matter' fluid,
which one can imagine as being
Cold Dark Matter,
or Hot Dark Matter (neutrinos),
or even high frequency gravitational waves.
The dark matter particles may be either massive or massless;
it does not change the character of the solutions.
The important thing is that
the dark matter passes freely through the baryons,
for the most part interacting with the baryons only by gravity,
although we do consider what happens if the dark matter has
a finite cross-section for absorption by the baryons.
As expected,
relativistic counter-streaming between baryons
and dark matter leads to mass inflation near the inner horizon.

Analytic and numerical work
on spherically symmetric collapse
and mass inflation
has commonly modeled the fluid accreted by a black hole
as a massless scalar field,
usually uncharged
\cite{Christodoulou86,GP87,GG93,BS95,Brady95,Burko97,BO98,Burko99,
HO01,Burko02a,Burko03,MG03,Dafermos04,HKN05},
but also charged
\cite{HP96,HP98b,HP99,SP01,OP03,Dafermos03}.
A key property of a massless scalar field
is that it allows waves to counter-stream relativistically through each other,
which allows mass inflation to occur.

In the present paper
we choose to adopt a somewhat different approach.
A driving motivation for this paper and its companion, Paper~1,
was to enquire what happens inside an astronomically realistic black hole,
and it has seemed to us that a mix of baryons and dark matter
might offer a more realistic description than a massless scalar field.
Besides this,
we wanted the freedom to explore what happens if the parameters of the model
are changed:
the electrical conductivity of the baryons (considered in Paper~1),
massive versus massless dark matter,
the ratio of accreted dark matter to baryonic density,
or the interaction cross-section between dark matter and baryons.

As in Paper~1,
our goal is not so much to study the formation of a black hole by
gravitational collapse,
but rather to explore the interior structure of black holes
after their formation.
We have in mind the situation of a realistic astronomical black hole,
perhaps stellar-sized, perhaps supermassive,
which is being fed by accretion of matter.

In a supermassive black hole especially,
the mass that the black hole acquires by gradual accretion
over the course of millions or billions of years
can greatly exceed the mass of the seed black hole,
formed perhaps from the collapse of the core of a massive star.
Under such circumstances it is reasonable to expect that the bulk
of the interior structure of the black hole is determined by the accretion
history, rather than by the details of the initial collapse event.

The situation considered in the present paper
is in a sense the opposite to that considered by
Poisson \& Israel (1990) \cite{PI90}
and others
\cite{Ori91,BDIM94,BS95,BO98,Dafermos03,Dafermos04},
who supposed that ingoing radiation or scalar field falling into a pre-existing
charged black hole would encounter a Price tail
\cite{Price72,HP98a,Dafermos03,Dafermos04}
of outgoing radiation
generated during gravitational collapse.
Because the black holes considered in the present paper are assumed
to accrete self-similarly into the indefinite future,
the theorems recently proven by Dafermos (2003)
\cite{Dafermos03}
concerning the collapse of isolated self-gravitating systems
and the decay of Price tails do not apply.
In the present paper,
accreted charged baryons
are repelled by the charge of the black hole self-consistently
produced by previously accreted baryons,
and naturally become outgoing.
Any outgoing Price tail of radiation generated by gravitational collapse
would soon be overwhelmed by the accreted charged baryons.
For example,
a black hole destined to become a quasar might radiate a fraction
of a solar mass in a Price tail at the time it forms from the
core collapse of a massive star,
but may subsequently accrete $10^8$ solar masses or more of baryons.

Since the bulk of the fluid inside the black hole is naturally outgoing,
to produce mass inflation the black hole must be allowed to accrete a
(small amount of) fluid which remains ingoing,
and which streams more or less freely through the outgoing baryons.
Dark matter fits the description nicely.

Many previous papers have found that the collapse of a massless scalar
field into a charged black hole produces
not only a strong spacelike singularity at zero radius
but also a weak null singularity at finite radius along the Cauchy horizon
\cite{Ori91,BDIM94,BS95,Burko97,BO98,HP98b,HP99,Ori99,HKN05},
as already anticipated by Poisson \& Israel (1990) \cite{PI90}.
A null singularity does not form in the similarity solutions.
Why this should be is discussed in \S\ref{nonull}.

An important question considered in the present paper
is what happens if dark matter and baryons have a finite
cross-section for interaction at high energy.
For mass inflation to persist,
and in particular for a null singularity to form on the Cauchy horizon,
it is necessary that counter-streaming ingoing and outgoing fluids
accelerate to arbitrarily close to the speed of light
(or become arbitrarily highly blueshifted, for massless streams)
relative to each other.
This raises the question of whether it is physically realistic
to allow counter-streaming at arbitrarily large Lorentz factors.
To address this question,
we explore the consequences of allowing the dark matter
to have a non-zero cross-section for absorption by baryons
at high collision energy.

The structure of this paper is as follows.
Section~\ref{Equations},
which is a follow-on to \S{}II of Paper~1,
presents the general relativistic equations governing
the interior and exterior structure of a spherically symmetric black hole
that accretes dark matter in addition to charged baryons.
Section~\ref{Similaritysolutions}
brings in the hypothesis of self-similarity,
and sets out the equations that follow from that hypothesis,
generalizing \S{}III of Paper~1.
Section~\ref{ResultsbaryonsDM}
gives results for self-similar black holes
accreting baryons and dark matter.
Section~\ref{Inflation}
discusses the physical question of why mass inflation occurs.
Section~\ref{Appearance}
addresses the question of what it would actually look like if
you fell into one of the black holes described herein.
Finally,
section~\ref{Summary}
summarizes the findings of this paper.

%Thus the present paper may offer a clue
%to the mystery of what happens beyond a singular Cauchy horizon
%\cite{Burko02b}.

\section{Equations}
\label{Equations}

This section
presents the general relativistic equations governing
a spherically symmetric black hole
accreting almost non-interacting dark matter
in addition to charged, electrically conducting baryons.
The required formalism has already been developed
for the most part in \S{}II of Paper~1 \cite{Paper1},
to which the reader is referred for notation and further detail.

\subsection{Frames}
\label{frames}

It proves convenient to work in the rest frame of the baryons,
with the dark matter streaming
at 4-velocity $\utet_d^m$
relative to the tetrad frame of the baryons.
To be consistent with spherical symmetry,
the dark matter must stream radially,
so the non-vanishing components of the dark matter 4-velocity $\utet_d^m$ are
the time and radial components
\begin{equation}
  \utet_d^0 = \utet_d^t
  \ , \quad
  \utet_d^i = \utet_d^r \, \hat x_i
  \ .
\end{equation}
Without loss of generality,
the rest mass $\mu$ of the dark matter particle can be taken
to be either $0$ or $1$,
depending on whether the particle is massless or massive.
Thus
\begin{equation}
  ( \utet_d^t )^2 - ( \utet_d^r )^2 = \mu^2
  \ , \quad
  \mu = \left\{
  \begin{array}{ll}
    0 & \mbox{massless} \\
    1 & \mbox{massive}
  \end{array}
  \right.
  \ .
\end{equation}
Below we will generally present arguments
as if the dark matter particle were massive.
The case of a massless particle follows from letting the massive
particle approach the speed of light
and simultaneously letting its rest mass go to zero.
Appropriate factors of rest mass $\mu$ are included in the formulae below.

Adapted to the dark matter frame is an associated tetrad frame,
with inertial axes $\gammavec_{d,m}$,
corresponding vierbein coefficients
$\alpha_d$, $\beta_d$, and $\gamma_d$,
equation~(6) of Paper~1,
and corresponding time coordinate $t_d$.
The dark matter time coordinate $t_d$
%adapted to the dark matter frame
differs from the baryonic time coordinate $t$
%adapted to the baryonic frame
because the gauge of time is being chosen, equation~(5) of Paper~1,
so that the proper radial derivative of time is zero;
that is,
the dark matter time coordinate $t_d$ is arranged to satisfy
$\partial_{d,r} t_d = 0$,
whereas the baryonic time coordinate $t$ satisfies
$\partial_r t = 0$.

The locally inertial axes $\gammavec_{d,m}$
in the rest frame of the dark matter
are related to the locally inertial axes $\gammavec_m$
in the baryonic frame
by a Lorentz boost at 4-velocity $\utet_d^m$.
%Note that this is a true Lorentz boost only if the 4-velocity
%$\utet_d^m$ has unit magnitude, $\mu^2 = 1$,
%but the formulae below remain valid also
%for massless dark matter, $\mu^2 = 0$.
It follows that the directed derivatives
in the dark matter frame,
$\partial_{d,t} \equiv \gammavec_{d,t} \cdot \partialvec$
and
$\partial_{d,r} \equiv \gammavec_{d,r} \cdot \partialvec$,
are equal to the directed derivatives in the baryonic frame
Lorentz boosted by 4-velocity $\utet_d^m$:
\begin{equation}
\label{ddtr}
  \partial_{d,t} = \utet_d^t \partial_t + \utet_d^r \partial_r
  \ , \quad
  \partial_{d,r} = \utet_d^r \partial_t + \utet_d^t \partial_r
\end{equation}
or more explicitly
\begin{eqnarray}
\label{dd}
&&\!\!\!\!\!\!
  \alpha_d \! \left. {\partial \over \partial t_d} \right|_r \!
  +
  \beta_d \! \left. {\partial \over \partial r} \right|_{t_d} \!
  =
  \utet_d^t \left(
    \alpha \! \left. {\partial \over \partial t} \right|_r \!
    +
    \beta \! \left. {\partial \over \partial r} \right|_{t}
  \right)
  + \utet_d^r \, \gamma \! \left. {\partial \over \partial r} \right|_{t}
\nonumber
\\
&&\!\!\!\!\!\!
  \gamma_d \! \left. {\partial \over \partial r} \right|_{t_d} \!
  =
  \utet_d^t \, \gamma \! \left. {\partial \over \partial r} \right|_{t}
  + \utet_d^r \left(
    \alpha \! \left. {\partial \over \partial t} \right|_r \!
    +
    \beta \! \left. {\partial \over \partial r} \right|_{t}
  \right)
  \ .
\end{eqnarray}

Applying equations~(\ref{dd}) to the radial coordinate $r$
shows that
$\beta_d$
and
$\gamma_d$
are equal to $\beta$ and $\gamma$
Lorentz boosted by 4-velocity $\utet_d^m$
\begin{equation}
\label{bgud}
  \beta_d = \beta \utet_d^t + \gamma \utet_d^r
  \ , \quad
  \gamma_d = \gamma \utet_d^t + \beta \utet_d^r
  \ .
\end{equation}
Equations~(\ref{bgud}) reflect the fact that
$(\beta, \gamma) = (\partial_t r, \partial_r r)$
form the time and radial components of a covariant 4-vector,
the radial 4-gradient,
as already remarked in equation~(12) of Paper~1.
%Note that the transformation~(\ref{bgud}) is, like (\ref{ddtr}),
%a true Lorentz transformation only if the 4-velocity $\utet_d^m$
%has unit magnitude, which is true only if $\mu^2 = 1$.
The magnitude squared of the covariant 4-vector
$(\beta_d, \gamma_d)$ is
\begin{equation}
\label{bgd}
  \beta_d^2 - \gamma_d^2 = \mu^2 ( \beta^2 - \gamma^2 )
\end{equation}
which is null for massless dark matter, $\mu^2 = 0$,
or the same as the magnitude squared of $(\beta, \gamma)$
for massive dark matter, $\mu^2 = 1$.

Similarly,
applying equations~(\ref{dd}) to the time coordinate $t_d$
yields the result that
\begin{equation}
\label{dtd}
  \left. {\partial t_d \over \partial t} \right|_{r}
  = {\alpha_d \gamma_d \over \alpha \gamma}
  \ , \quad
  \left. {\partial t_d \over \partial r} \right|_{t}
  = - {\alpha_d \utet_d^r \over \gamma}
  \ .
\end{equation}

For self-similar solutions,
equations~(\ref{dtd})
translate into a second explicit relation
[the first being equations~(\ref{bgud})]
between the dark matter and baryon vierbein coefficients.
The second relation, equation~(\ref{xid}) below,
is most transparently derived not directly
from equations~(\ref{dtd}),
but rather from the transformation of the homothetic 4-vector
introduced in \S{}III\,B of Paper~1 \cite{Paper1}.
%Derivation of this relation is postponed to \S\ref{homothetic},
%where the homothetic 4-vector $\xivec$ is introduced.
%With the homothetic vector in hand,
%the desired relation between vierbein coefficients reduces to the
%statement that the homothetic 4-vector in the dark matter frame
%is a Lorentz boosted version of the homothetic 4-vector in the baryonic frame,
%equation~(\ref{xid}).
The components $\xi_d^m$ of the homothetic vector
in the tetrad frame of the dark matter are equal to those in the baryonic frame
Lorentz boosted by the 4-velocity $\utet_d^m$ of the dark matter
relative to the baryons
\begin{equation}
\label{xid}
  \xi_d^t
  =
  {1 \over \eta_d}
  =
  \xi^t \utet_d^t - \xi^r \utet_d^r
  \ , \quad
  \xi_d^r
  =
  {V_d \over \eta_d}
  =
  \xi^r \utet_d^t - \xi^t \utet_d^r
  \ .
\end{equation}
%Equations~(\ref{xid})
%constitute the explicit second relation between the baryonic and dark matter
%vierbein coefficients referred to at the end of \S\ref{frames}.
The magnitude squared of the homothetic vector in the dark matter frame is
\begin{equation}
  ( \xi_d^t )^2 - ( \xi_d^r )^2
  =
  \mu^2 H
\end{equation}
which is null for massless dark matter, $\mu^2 = 0$,
or equal to the homothetic scalar $H$,
equation~(58) of Paper~1, for massive dark matter, $\mu^2 = 1$.

\subsection{Einstein equations}

In \S{}II\,A of Paper~1,
the tetrad formalism for spherically symmetric geometry was set up
and the Einstein and Weyl tensors derived,
and in \S{}II\,B of Paper~1,
the resulting Einstein equations were written down for the
case where the tetrad frame was taken to be the center of mass frame,
defined to the frame where the momentum density vanishes.
In the present subsection
the Einstein equations are given for a radially moving tetrad frame,
such as the baryonic frame,
which is not necessarily the center of mass frame.
The Einstein equations~(\ref{einsteineqsd}) in the moving frame are
not as physically transparent as those, equations~(34) of Paper~1,
in the center of mass frame.

The most general form of the energy-momentum tensor $T_{mn}$ consistent
with spherical symmetry is
\begin{eqnarray}
&&
  T_{00} = T^{tt}
  \ , \quad
  T_{0i} = - T^{tr} \, \hat x_i
  \ ,
\nonumber
\\
&&
  T_{ij}
  =
  T^{rr} \, \hat x_i \hat x_j
  + T^{\perpperp} \left( \delta_{ij} - \hat x_i \hat x_j \right)
  \ .
\end{eqnarray}
Substituting the Einstein tensor, equation~(26) of Paper~1,
into Einstein's equations
$G_{mn} = 8 \pi T_{mn}$
implies that
the quantities $F$, $R$, $P$, and $S$
defined by equations~(25) of Paper~1 are
\begin{eqnarray}
&&
  R = 4 \pi T^{tt}
  \ , \quad
  F = 4 \pi T^{tr}
  \ ,
\nonumber
\\
&&
\label{FRPSeq}
  P = 4 \pi T^{rr}
  \ , \quad
  S = 4 \pi ( T^{\perpperp} - T^{rr} )
  \ .
\end{eqnarray}
The Einstein equations thus become
\begin{subequations}
\label{einsteineqsd}
\begin{eqnarray}
\label{massdensityd}
&
\displaystyle{
  \partial_r M - 4 \pi r^2 \left( \gamma \, T^{tt} + \beta \, T^{tr} \right)
  = 0
  \ ,
}
&
\\
\label{bernoullid}
&
\displaystyle{
  \partial_t \gamma - \beta g - 4 \pi r T^{tr}
  = 0
  \ ,
}
&
\\
\label{accelerationd}
&
\displaystyle{
  \partial_t \beta - \gamma g + {M \over r^2} + 4 \pi r T^{rr}
  = 0
  \ ,
}
&
%\\
%\label{eulerd}
%&
%\displaystyle{
%  \partial_t T^{tr} + 2 T^{tr} \left( {\partial \beta \over \partial r} + {\beta \over r} - {r F \over \gamma} \right)
%  \qquad\qquad\qquad
%}
%&
%\\
%\nonumber
%&
%\displaystyle{
%  + \,
%  \partial_r T^{rr} - {2 \gamma \over r} \left( T^{\perpperp} - T^{rr} \right) + \left( T^{tt} + T^{rr} \right) g
%  = 0
%}
%&
\\
\label{eulerd}
&
\displaystyle{
  \partial_t T^{tr}
  + \partial_r T^{rr}
  - {2 \gamma \over r} \left( T^{\perpperp} - T^{rr} \right)
  \qquad
}
&
\nonumber
\\
&
\displaystyle{
  + \, 2 \Bigl( {\beta \over r} + h \Bigr) T^{tr}
  + g \left( T^{tt} + T^{rr} \right)
  = 0
  \ .
}
&
\end{eqnarray}
\end{subequations}
Equation~(\ref{bernoullid})
implies that the quantity $h$,
defined by equation~(22) of Paper~1,
satisfies
\begin{equation}
  h = {\partial \beta \over \partial r} - {4 \pi r T^{tr} \over \gamma}
  \ .
\end{equation}
The result of taking $\beta$ times equation~(\ref{accelerationd})
minus $\gamma$ times equation~(\ref{bernoullid}) is
the first law of thermodynamics for the combined baryonic
and dark matter fluid
\begin{equation}
\label{energyconservationd}
  \partial_t M + 4 \pi r^2 \left( \gamma \, T^{tr} + \beta \, T^{rr} \right)
  = 0
  \ .
\end{equation}
Einstein's equations automatically incorporate
conservation of energy-momentum,
as expressed by the vanishing of the covariant derivative of the
energy-momentum tensor,
$D_m T^{mn} = 0$.
The time ($n = 0$) component of the energy-momentum conservation equation
gives the energy conservation equation
%\begin{eqnarray}
%\label{dtrhod}
%&&
%  \partial_t T^{tt}
%  + \partial_r T^{tr} + 2 \left( {\gamma \over r} + g \right) T^{tr}
%  + {2 \beta \over r} \left( T^{\perpperp} - T^{rr} \right)
%\nonumber
%\\
%&&
%  + \, \left( T^{tt} + T^{rr} \right) \left( {\partial \beta \over \partial r} + {2 \beta \over r} - {r F \over \gamma} \right)
%  = 0
%\end{eqnarray}
\begin{eqnarray}
\label{dtrhod}
&&
  \partial_t T^{tt}
  + \partial_r T^{tr}
  + {2 \beta \over r} \left( T^{\perpperp} - T^{rr} \right)
\nonumber
\\
&&
  + \, 2 \left( {\gamma \over r} + g \right) T^{tr}
  + \Bigl( {2 \beta \over r} + h \Bigr) \left( T^{tt} + T^{rr} \right)
  = 0
  \quad
\end{eqnarray}
while the spatial components ($n = 1,2,3$)
of the energy-momentum conservation equation
give a momentum conservation equation
which reduces precisely to the Euler
equation~(\ref{eulerd}).

\subsection{Dark matter}

We take the dark matter (subscripted $d$)
to be a pressureless fluid (dust, either massive or massless)
that free falls radially into the black hole.
The energy-momentum tensor of the dark matter is thus
\begin{equation}
\label{Td}
  T_d^{mn} = \rho_d \, \utet_d^m \utet_d^n
\end{equation}
whose non-zero components are
\begin{equation}
  T_d^{tt} = \rho_d \, (\utet_d^t)^2
  \ , \quad
  T_d^{tr} = \rho_d \, \utet_d^t \utet_d^r
  \ , \quad
  T_d^{rr} = \rho_d \, (\utet_d^r)^2
  \ .
\end{equation}

We assume that the dark matter is almost non-interacting,
for the most part streaming freely through the baryons,
but we wish to consider the possibility, \S\ref{interaction},
that dark matter particles interact with baryons
when they pass through each other at sufficiently high energy.
For simplicity,
we assume that dark matter particles that interact with the baryons
are simply absorbed into the baryonic fluid,
adding their energy and momentum into the baryons,
which retain their isotropic pressure and relativistic equation of state.
If the rate,
the mass per unit volume per unit time,
at which dark matter is absorbed into the baryonic fluid is denoted
$\dot\rho_d$,
in the frame of reference of the dark matter,
then the equations of energy-momentum conservation for the dark matter are
\begin{equation}
\label{DTd}
  D_m T_d^{mn} = - \dot\rho_d \, \utet_d^n
  \ .
\end{equation}
For the energy-momentum tensor of equation~(\ref{Td}),
the energy-momentum conservation equations~(\ref{DTd})
are equivalent to two equations,
first, the equation describing unaccelerated free-fall of the dark matter,
\begin{equation}
\label{Dud}
  {D \utet_d^n \over D \tau_d} = 0
\end{equation}
where $D / D \tau_d = \utet_d^m D_m$,
and second, the equation for conservation of dark matter rest mass
\begin{equation}
\label{Drhod}
  D_m \left( \rho_d \, \utet_d^m \right) = - \dot\rho_d
  \ .
\end{equation}
The free-fall equation~(\ref{Dud}) gives
%\begin{equation}
%\label{dud}
%  \partial_{d,t} \utet_d^r
%%  \left( \utet_d^r \partial_t + \utet_d^t \partial_r \right) \utet_d^r
%  + g ( \utet_d^t )^2
%  + \left( {\partial \beta \over \partial r} - {r F \over \gamma} \right)
%  \utet_d^t \utet_d^r
%  = 0
%\end{equation}
\begin{equation}
\label{dud}
  \partial_{d,t} \utet_d^r
%  \left( \utet_d^r \partial_t + \utet_d^t \partial_r \right) \utet_d^r
  + \utet_d^t \left( g \utet_d^t + h \utet_d^r \right)
  = 0
\end{equation}
while equation~(\ref{Drhod}) for conservation of dark matter rest mass is
%\begin{eqnarray}
%  &&
%  \left( \partial_t + {\partial \beta \over \partial r} + {2 \beta \over r} - {r F \over \gamma} \right)
%  \left( \rho_d \utet_d^t \right)
%\nonumber
%\\
%  &&
%  + \,
%  \left( \partial_r + {2 \gamma \over r} + g \right)
%  \left( \rho_d \utet_d^r \right)
%  =
%  - \dot\rho_d
%  \ .
%\end{eqnarray}
\begin{equation}
  \Bigl( \partial_t + {2 \beta \over r} + h \Bigr)
  \left( \rho_d \utet_d^t \right)
  +
  \Bigl( \partial_r + {2 \gamma \over r} + g \Bigr)
  \left( \rho_d \utet_d^r \right)
  =
  - \dot\rho_d
  \ .
\end{equation}
%\begin{equation}
%  \partial_{d,t} M_d = \dot M_d
%  \ , \quad
%  \partial_{d,r} M_d = 4 \pi r^2 \rho_d
%\end{equation}

\subsection{Baryons and electric field}

The baryons (subscripted $b$)
and the electric field (subscripted $e$) are coupled.
As remarked in \S\ref{frames},
it is simplest to work in the rest frame of the baryons,
where the momentum density of baryons is zero $T_b^{tr} = 0$.
The energy-momentum tensor of the baryons
in the tetrad frame of the baryons is diagonal, with non-zero components
(cf.\ \S{}II\,C of Paper~1)
\begin{equation}
  T_b^{tt} = \rho_b
  \ , \quad
  T_b^{rr} = T_b^{\perpperp} = p_b
\end{equation}
in which the density $\rho_b$ and pressure $p_b$
are assumed to be related by a relativistic equation of state
\begin{equation}
\label{w}
  p_b = w \rho_b
  \ , \quad
  w = \frac{1}{3}
  \ .
\end{equation}
The energy-momentum tensor of the electric field is similarly diagonal,
with non-zero components
(cf.\ \S{}II\,D of Paper~1)
\begin{equation}
  T_e^{tt} = - T_e^{rr} = T_e^{\perpperp} = \rho_e
\end{equation}
where the electric energy density is
$\rho_e = Q^2 / ( 8 \pi r^4 )$,
with $Q$ the charge interior to radius $r$.

Together, the coupled baryons and electric field
satisfy the energy-momentum conservation equation
\begin{equation}
\label{DTbe}
  D_m \left( T_b^{mn} + T_e^{mn} \right)
  = \dot\rho_d \utet_d^n
\end{equation}
the right hand side of which represents
the energy-momentum dumped into the baryonic fluid
as the result of dark matter being absorbed by baryons.
The time ($n = 0$) and space ($n = 1,2,3$)
components of
%the energy-momentum conservation
equation~(\ref{DTbe})
yield the energy and momentum conservation equations
for the baryons coupled to the electric field
%\begin{align}
%\label{dtrhobe}
%  \partial_t \left( \rho_b + \rho_e \right)
%  + {4 \beta \over r} \rho_e
%  + \left( \rho_b + p_b \right) \biggl( & {\partial \beta \over \partial r} + {2 \beta \over r} - {r F \over \gamma} \biggr)
%\nonumber
%\\
%  &
%  = \dot\rho_d \, \utet_d^t
%\\
%  \partial_r \left( p_b - \rho_e \right)
%  - {4 \gamma \over r} \rho_e
%  + \left( \rho_b + p_b \right) g
%  &
%  = \dot\rho_d \, \utet_d^r
%\end{align}
\begin{align}
\label{dtrhobe}
  &
  \partial_t \left( \rho_b + \rho_e \right)
  + {4 \beta \over r} \rho_e
  + \left( \rho_b + p_b \right) \left( {2 \beta \over r} + h \right)
  = \dot\rho_d \, \utet_d^t
\\
  &
  \partial_r \left( p_b - \rho_e \right)
  - {4 \gamma \over r} \rho_e
  + \left( \rho_b + p_b \right) g
  = \dot\rho_d \, \utet_d^r
  \ .
\end{align}

%\begin{eqnarray}
%\label{dtrhobe}
%  &&
%  \partial_t \left( \rho_b + \rho_e \right)
%  + {4 \beta \over r} \rho_e
%  + \left( \rho_b + p_b \right) \left( {\partial \beta \over \partial r} + {2 \beta \over r} - {r F \over \gamma} \right)
%\nonumber
%\\
%  &&
%  \qquad\qquad\qquad\qquad\qquad\qquad\qquad\ \ 
%  = \dot\rho_d \, \utet_d^t
%\\
%  &&
%  \partial_r \left( p_b - \rho_e \right)
%  - {4 \gamma \over r} \rho_e
%  + \left( \rho_b + p_b \right) g
%  = \dot\rho_d \, \utet_d^r
%\end{eqnarray}
%
%\begin{eqnarray}
%\label{dtrhobe}
%  &&
%\!\!\!\!\!\!\!\!\!\!\!\!\!\!\!\!\!\!\!\!
%  \partial_t T_{be}^{tt}
%  + {2 \beta \over r} \left( T_{be}^{\perpperp} - T_{be}^{rr} \right)
%\nonumber
%\\
%  &&
%\!\!\!\!\!\!\!\!\!\!\!\!\!\!\!\!\!\!\!\!
%  + \, \left( T_{be}^{tt} + T_{be}^{rr} \right) \left( {\partial \beta \over \partial r} + {2 \beta \over r} - {r F \over \gamma} \right)
%  = \dot\rho_d \, \utet_d^t
%\\
%  &&
%\!\!\!\!\!\!\!\!\!\!\!\!\!\!\!\!\!\!\!\!
%  \partial_r T_{be}^{rr}
%  - {2 \gamma \over r} \left( T_{be}^{\perpperp}
%  - T_{be}^{rr} \right) + \left( T_{be}^{tt} + T_{be}^{rr} \right) g
%  = \dot\rho_d \, \utet_d^r
%\end{eqnarray}

\subsection{Interaction between dark matter and baryons}
\label{interaction}

We assume that the rate $\dot\rho_d$
at which dark matter is absorbed into baryons
is proportional to the density $\rho_d$ of dark matter particles,
multiplied by an absorption rate per particle $\sigma_d$
\begin{equation}
\label{rhoddot}
  \dot\rho_d = \rho_d \sigma_d
  \ .
\end{equation}
In general the rate $\sigma_d$
will be an integral over collision energy
(in the dark matter frame)
of the number density of baryons
times the absorption cross-section
times the collision velocity.
We treat the absorption rate $\sigma_d$
as a phenomenological quantity,
the aim being to explore what happens
as the properties of the absorption rate are varied.
If the absorption rate $\sigma_d$ is assumed
to be some function of the baryonic density $\rho_b$
and of the 4-velocity $\utet_d^m$
between the dark matter and the baryons,
then self-similarity requires
that the absorption rate be proportional to the square root of
the baryonic density,
similarly to the electrical conductivity,
equation~(48) of Paper~1,
and otherwise to be some arbitrary function of the 4-velocity.
Thus the absorption rate $\sigma_d$ is taken to be
\begin{equation}
\label{sigmad}
  \sigma_d
  =
  \kappa_d
  \left( 4 \pi \rho_b \right)^{1/2}
\end{equation}
where $\kappa_d$
is a phenomenological dimensionless rate coefficient,
which in general could be some arbitrary function of the 4-velocity
$\utet_d^m$
between dark matter and baryons.
The factor of $4 \pi$ in equation~(\ref{sigmad})
is introduced to simplify the corresponding self-similar equation~(\ref{sd}).

\section{Similarity solutions}
\label{Similaritysolutions}

\subsection{Similarity hypothesis}

As noted in Paper~1,
dimensional analysis reveals the following quantities to be dimensionless
[the following equation repeats equations~(49) and (50) of Paper~1]
\begin{eqnarray}
\label{dimensionless}
&
\nonumber
\displaystyle{
  \eta \equiv {\alpha r \over t}
  \ , \quad
  \beta
  \ , \quad
  \gamma
  \ , \quad
  {M \over r}
  \ , \quad
  {Q \over r}
  \ ,
}
&
\\
\nonumber
&
\displaystyle{
  y \equiv g r
  \ , \quad
  z \equiv 4 \pi r^2 \rho_b
  \ , \quad
  z_q \equiv 4 \pi r^2 q
  \ , \quad
  s \equiv 4 \pi r \sigma
  \ ,
}
&
\\
&
\displaystyle{
\label{ze}
  z_e \equiv 4 \pi r^2 \rho_e = {Q^2 \over 2 r^2}
  \ ,
}
\end{eqnarray}
where the dimensionless conductivity $s$ is
\begin{equation}
\label{sigmasim}
  s = \kappa z^{1/2}
  \ .
\end{equation}
With dark matter adjoined,
dimensional analysis of the previous equations,
combined with equation~(\ref{Drhod})
for conservation of dark matter energy-momentum,
%equation~(\ref{Td}),
and equation~(\ref{rhoddot})
for the dark matter absorption rate,
shows that the following dark matter quantities are likewise dimensionless:
\begin{eqnarray}
\label{dimensionlessd}
&
\displaystyle{
  \eta_d \equiv {\alpha_d r \over t_d}
  \ , \quad
  \beta_d
  \ , \quad
  \gamma_d
  \ , \quad
  \utet_d^m
  \ ,
}
&
\\
\nonumber
&
\displaystyle{
  z_d \equiv 4 \pi r^2 \rho_d
  \ , \quad
  s_d \equiv r \sigma_d
  \ ,
}
&
\end{eqnarray}
where the dimensionless dark matter absorption rate $s_d$ is,
for the phenomenological dark matter absorption rate $\sigma_d$ given by
equation~(\ref{sigmad}),
\begin{equation}
\label{sd}
  s_d
  =
  \kappa_d
  z^{1/2}
  \ .
\end{equation}

The dimensionless variables of equations~(\ref{dimensionless})
and (\ref{dimensionlessd})
form a (more than) complete set for the problem at hand,
and it follows
\cite{CC99}
that the spherically symmetric Einstein-Maxwell and subsidiary equations
admit similarity solutions in which the dimensionless variables
are all functions of a single dimensionless variable.

As shown in Paper~1, equation~(53),
the proper radial velocity $V$ of the similarity frame
relative to the baryonic tetrad frame is
\begin{equation}
\label{V}
  V
  = {\eta - \beta \over \gamma}
  \ .
\end{equation}
The radial velocity $V_d$ of the similarity frame
relative to the dark matter tetrad frame is similarly
\begin{equation}
\label{Vd}
  V_d
  = {\eta_d - \beta_d \over \gamma_d}
  \ .
\end{equation}

\subsection{Integrals of the similarity equations}

The ordinary differential equations
determining the self-similar evolution of the baryonic and dark matter fluids
admit four integrals,
of which three are generalizations of the three integrals
given in \S{}III\,F of Paper~1,
and the fourth is an integral for the dark matter.

The first integral follows from equation~(87) of Paper~1,
which here implies
\begin{equation}
\label{Mrd}
  {M \over r}
  =
%  {z ( \gamma V - w \beta ) \over \eta}
  z ( \gamma \xi^r - w \beta \xi^t )
  +
  z_e
  +
%  {z_d \gamma_d V_d \over \eta_d}
  z_d \gamma_d \xi_d^r
  \ .
\end{equation}
Equation~(\ref{Mrd}) differs from the corresponding equation~(88) of Paper~1
by the addition of the last, dark matter, term on the right hand side.
As in Paper~1,
we use equation~(\ref{Mrd})
not as one of the evolutionary equations,
but rather as a check on the accuracy of the integration.

The second integral of the similarity equation
is unchanged from Paper~1.
The integral follows from equation~(89) of Paper~1,
which yields an equation for the dimensionless charge density
$z_q \equiv 4 \pi r^2 q$
[the following repeats equation~(90) of Paper~1]
\begin{equation}
\label{zq}
  z_q
  =
%  {Q ( \eta + s ) \over r V}
  {Q ( 1 + s \xi^t ) \over r \xi^r}
  \ .
\end{equation}

The third integral of the similarity equations follows from
equation~(91) of Paper~1,
which here yields a revised equation for the dimensionless proper acceleration
$y \equiv g r$
%\begin{equation}
%\label{y}
%  y
%  =
%  {V \left\{
%  \begin{array}{l}
%%    w z \left[ 2 \eta - 2 (1 {+} w ) \beta - (1 {+} w)^2 z / \eta \right]
%%    \\
%%    + \, 2 z_e \left[ \eta + (1 {+} w) s \right]
%    2 w z ( \gamma V - w \beta )
%    + 2 z_e \left[ \eta + (1 {+} w) s \right]
%    \\
%    - \, w (1 {+} w) z (1 {+} w) z / \eta
%  \end{array}
%  \right\}
%  \over
%  (1 {+} w) z ( V^2 - w )}
%  \ .
%\end{equation}
\begin{equation}
\label{yd}
  y
  =
  {\left\{
  \begin{array}{l}
    2 w \xi^r M / r
    \\
    + \, 2 z_e \xi^r \left[ (1 {-} w) + (1 {+} w) s \xi^t \right]
    \\
    + z_d \xi^r \left[ - \,2 w \gamma_d \xi_d^r
      + s_d \left( w \xi^t \utet_d^t + \xi^r \utet_d^r \right) \right]
    \\
    - \, w (1 {+} w) z \xi^t
    \left[ (1 {+} w) z \xi^t \xi^r + z_d \xi_d^t \xi_d^r \right]
  \end{array}
  \right\}
  \over
  (1 {+} w) z \left[ (\xi^r)^2 - w (\xi^t)^2 \right]}
  \ .
\end{equation}
%\begin{equation}
%\label{yd}
%  y
%  =
%  {V \left\{
%  \begin{array}{l}
%    2 w z ( \gamma V - w \beta )
%    \\
%    + \, 2 z_e \left[ \eta + (1 {+} w) s \right]
%    + z_d s_d \left( w \utet_d^t + V \utet_d^r \right)
%    \\
%    - \, w (1 {+} w) z \left[ (1 {+} w) z / \eta + k z_d / \eta_d \right]
%  \end{array}
%  \right\}
%  \over
%  (1 {+} w) z ( V^2 - w )}
%  \ .
%\end{equation}
Equation~(\ref{yd}) differs from the corresponding equation~(92) of Paper~1
by the addition of various dark matter terms
in the numerator on the right hand side.

A fourth integral follows from the geodesic integral of motion~(62) of Paper~1,
which applies to the freely-falling dark matter.
The homothetic momentum, equation~(63) of Paper~1,
of the freely-falling dark matter is
$\ucoord_{\lnt} = r \xi_m \utet_d^m = - r \xi_d^t = - t_d / \alpha_d$,
and the integral of motion~(62) of Paper~1 then shows that
\begin{equation}
\label{td}
  {t_d \over \alpha_d} = \tau_d
  \ .
\end{equation}
%The fact that only $t_d / \alpha_d$ is determined reflects the
%gauge ambiguity in the choice of time coordinate $t_d$.
%If one liked, one could fix the gauge in a natural way by setting
%$\alpha_d = 1$,
%equivalent (at least for massive dark matter particles)
%to synchronizing the dark matter coordinate time $t_d$
%to the proper time $\tau_d$ of freely-falling dark matter particles.
%However, there is no need to set the gauge,
%and equation~(\ref{td}) simply implies that
Equation~(\ref{td}) implies that
\begin{equation}
\label{xidt}
  \xi_d^t
  =
  {\tau_d \over r}
  \ .
\end{equation}
%\begin{equation}
%%  M^\prime_d
%%  \equiv
%  {M_d \over r_d} - {\dot M_d \over \eta_d}
%  =
%  {z_d V_d \over \eta_d}
%\end{equation}

\subsection{Similarity differential equations}

As in Paper~1,
we adopt as a suitable dimensionless integration variable
the dimensionless baryonic time parameter $x$
defined by equation~(93) of Paper~1.
The baryonic proper time $\tau$,
time coordinate $t$,
and radial coordinate $r$
evolve along the path of the baryonic fluid
according to the same equations as before,
equations~(94) of Paper~1,
which we repeat here for completeness:
\begin{subequations}
\begin{eqnarray}
\label{taux}
  {\dd \tau \over \dd x}
  &=&
  r
  \ ,
\\
\label{tx}
  {\dd \ln t \over \dd x}
  &=&
  \eta
  \ ,
\\
\label{rx}
  {\dd \ln r \over \dd x}
  &=&
  \beta
  \ .
\end{eqnarray}
\end{subequations}
Equation~(\ref{tx}) presumes that the gauge of baryonic time $t$ is
chosen in the natural way,
such that the units of time are the same as the units of radius,
so that $r / t$ is a dimensionless variable.

Similarly,
the dark matter proper time $\tau_d$,
time coordinate $t_d$,
and radial coordinate $r_d$
evolve along the path of the dark matter fluid as
\begin{subequations}
\begin{eqnarray}
\label{taudx}
  {\dd \tau_d \over \dd x}
  &=&
  {\mu^2 r_d \over k}
  \ ,
\\
\label{tdx}
  {\dd \ln t_d \over \dd x}
  &=&
  {\mu^2 \eta_d \over k}
  \ ,
\\
\label{rdx}
  {\dd \ln r_d \over \dd x}
  &=&
  {\beta_d \over k}
\end{eqnarray}
\end{subequations}
where
\begin{equation}
  k \equiv {\xi_d^r \over \xi^r}
  \ .
\end{equation}
Note that the radial coordinate $r_d$ in equation~(\ref{rdx})
is distinguished from the radial coordinate $r$ in equation~(\ref{rx}),
because the integration is along the path of the dark matter
in equation~(\ref{rdx})
as opposed to the path of the baryons in equation~(\ref{rx}).
With no charge to repel their infall,
dark matter particles fall in faster than baryons,
so that the dark matter radius $r_d$
is less than the baryonic radius $r$
at any given self-similar point inside the outer boundary.
For example,
if $r_d$ is half of $r$ at a given point,
it means that the black hole was twice as large when it accreted the dark matter
as it was when it accreted the baryons.

The differential equation~(\ref{tdx}) for the dark matter time coordinate $t_d$
again presumes that that the gauge of $t_d$ is
chosen such that $r_d / t_d$ is a dimensionless variable.
Equations~(\ref{taudx}) and (\ref{tdx}),
along with $\eta_d = 1 / \xi_d^t = r_d / \tau_d$ from equation~(\ref{xidt}),
imply that
$t_d \propto \tau_d$,
which together with equation~(\ref{td}) implies that $\alpha_d$ is a constant.
The constancy of $\alpha_d$
can also be regarded as following from the fact that the acceleration
$g_d \equiv - \partial_{d,r} \ln\alpha_d$, cf.\ equation~(21) of Paper~1,
vanishes for freely-falling dark matter,
$g_d = 0$;
the vanishing of the radial derivative of $\alpha_d$,
coupled with self-similarity,
implies that the total derivative $\dd \alpha_d / \dd x$ vanishes.
It is natural to adopt the gauge choice $\alpha_d = 1$,
so that the dark matter time coincides with dark matter proper time,
$t_d = \tau_d$.
However, the dark matter time $t_d$
(as distinct from dark matter proper time $\tau_d$)
is not actually used in this paper.

An overcomplete set of equations
[only three of the four equations~(\ref{dsim}) below are independent,
the four variables $\xi^t$, $\xi^r$, $\gamma$, and $\beta$
being related by
$\beta \xi^t + \gamma \xi^r = 1$,
equation~(\ref{xib}) below]
governing the self-similar evolution of the remaining variables is
[the following generalize equations~(95) of Paper~1]
\begin{subequations}
\label{dsim}
\begin{eqnarray}
\label{xitsim}
  {\dd \xi^t \over \dd x}
  &=&
  - \, y \xi^r + \gamma \xi^r
\\
\label{xirsim}
  {\dd \xi^r \over \dd x}
  &=&
  - \, y \xi^ t
  - \beta \xi^r
  + (1 {+} w) z \xi^t \xi^r
  + z_d \xi_d^t \xi_d^r
  \qquad
\\
\label{gamsim}
  {\dd \gamma \over \dd x}
  &=&
  \beta y + z_d \, \utet_d^t \utet_d^r
\\
\label{betsim}
  {\dd \beta \over \dd x}
  &=&
  \gamma y
  - (1 {+} w) z \gamma \xi^r
  - z_d \left[ \gamma_d \xi_d^r + (\utet_d^r)^2 \right]
\end{eqnarray}
\end{subequations}
together with
[the following generalize equations~(96) of Paper~1]
\begin{subequations}
\label{dsimz}
\begin{eqnarray}
  {\dd \ln [ r^{1 + 3 w} z ( \xi^r )^{1 + w} ] \over \dd x}
  &=&
  {2 z_e s \over z} + {z_d s_d \utet_d^t \over z}
  \qquad
\\
  {\dd \ln Q \over \dd x}
  &=&
  - \, s
\\
  {\dd \ln ( r_d z_d \xi_d^r ) \over \dd x}
  &=&
  - \, {s_d \over k}
  \ .
\end{eqnarray}
\end{subequations}
To maintain numerical precision,
it is important to avoid expressing small quantities
as differences of large quantities.
For example,
$\xi^t + \xi^r = ( 1 + V ) / \eta$
can be tiny near the Cauchy horizon, $V \approx -1$,
where mass inflation occurs,
though $\xi^t$ and $\xi^r$ are individually substantial.
A suitable choice of variables to integrate
%that works in all situations considered in this paper,
is
$\xi^t + \xi^r$,
$\beta - \gamma$,
$\gamma$,
$\tau_d$
and
$r_d$,
the last two giving $\xi_d^t$ according to equation~(\ref{xidt}).
Starting from these variables,
the following chain of equations yield
the remaining variables in a fashion that ensures numerical precision
[the following equations generalize equations~(97) of Paper~1]:
\begin{subequations}
\begin{eqnarray}
\label{xib}
&
\displaystyle{
  \xi^t - \xi^r
  =
  {2 - ( \beta + \gamma ) ( \xi^t + \xi^r ) \over \beta - \gamma}
}
&
\\
&
\displaystyle{
  H
  =
  ( \xi^t + \xi^r ) ( \xi^t - \xi^r )
}
&
\\
&
\displaystyle{
  {2 M \over r}
  =
  1 + ( \beta + \gamma ) ( \beta - \gamma )
}
&
\\
&
\displaystyle{
  \xi_d^r
  =
  \left[ (\xi_d^t)^2 - \mu^2 H \right]^{1/2}
}
\\
&
\displaystyle{
  \utet_d^t - \utet_d^r
  =
  {\xi_d^t + \xi_d^r \over \xi^t + \xi^r}
}
&
\\
&
\displaystyle{
  \utet_d^t + \utet_d^r
  =
  {\mu^2 \over \utet_d^t - \utet_d^r}
}
&
\\
&
\displaystyle{
  \beta_d \pm \gamma_d
  =
  \left( \beta \pm \gamma \right) \left( \utet_d^t \pm \utet_d^r \right)
  \ .
}
&
\end{eqnarray}
\end{subequations}

The differential equation for the variable $X$ used in ray-tracing,
\S{}III\,D of Paper~1,
remains unchanged from Paper~1
[the following repeats equation~(98) of Paper~1]
\begin{equation}
  {\dd X \over \dd x} = - \, \xi^r
  \ .
\end{equation}

During mass inflation,
the interior mass $M$
and (the absolute value of) the radial streaming 4-velocity $\utet_d^r$
increase exponentially,
while (the absolute value of) the homothetic scalar $H$ decreases exponentially.
The following differential equations,
which are consequences of the equations above,
are useful for characterizing mass inflation.
The interior mass $M$, which satisfies
$2 M / r - 1 = \beta^2 - \gamma^2$,
equation~(24) of Paper~1,
evolves as
\begin{equation}
\label{dlnM}
  {\dd \ln \left| 2 M / r - 1 \right| \over \dd x}
  =
  - \,
  8 \pi r^2 T_\gamma^{tr} \xi^r
\end{equation}
where $T_\gamma^{tr}$
is the proper momentum density relative to the $\gamma = 0$ frame
\begin{equation}
\label{Tgamma}
  4 \pi r^2 T_\gamma^{tr}
  =
  {(1 {+} w) z \beta \gamma + z_d \beta_d \gamma_d
  \over 2 M / r - 1}
  \ .
\end{equation}
The homothetic scalar
$H \equiv (\xi^t)^2 - (\xi^r)^2$,
equation~(58) of Paper~1,
evolves as
\begin{equation}
\label{dlnH}
  {\dd \ln \left| H \right| \over \dd x}
  =
  \left( {2 ( \gamma \xi^t + \beta \xi^r ) \over H}
    + 8 \pi r^2 T_\xi^{tr} \right)
  \xi^r
\end{equation}
where $T_\xi^{tr}$
is the momentum density relative to the no-going $\xi^t = 0$ frame,
the frame of reference at the border between ingoing (positive $\xi^t$)
and outgoing (negative $\xi^t$),
\begin{equation}
\label{Txi}
  4 \pi r^2 T_\xi^{tr}
  =
  {(1 {+} w) z \xi^t \xi^r + z_d \xi_d^t \xi_d^r
  \over - H}
  \ .
\end{equation}
The 4-velocity $\utet_d^r$ of the dark matter through the baryons
evolves as
\begin{equation}
\label{dlnu}
  {\dd \ln \left| \utet_d^r \right| \over \dd x}
  =
  \left( - \, {y \over \utet_d^r}
    + {4 \pi r^2 H T_\xi^{tr} \over \xi_d^r} \right)
  \utet_d^t
  \ .
\end{equation}

\subsection{Boundary conditions at the outer sonic point}
\label{boundaryconditions}

As in Paper~1,
the boundary conditions of the calculation are set at an outer boundary,
taken to be a regular sonic point,
outside the outer horizon of the black hole,
where the infalling baryonic fluid transitions smoothly
from subsonic to supersonic.

Two boundary conditions at the outer sonic point
are carried over from
\S{}III\,H of Paper~1,
namely the accretion rate $\eta_s$,
and the charge-to-mass ratio $Q/\Mc$ of the black hole.

Dark matter adds two more boundary conditions,
which set the velocity and density of dark matter at the outer sonic point.
If there were no mass or charge outside the outer sonic point,
then dark matter that free-falls from zero velocity at infinity
would have $\gamma_d = 1$,
and we adopt this value as a natural choice
for setting the velocity of the dark matter at the outer boundary:
\begin{equation}
\label{gammadsonic}
  \gamma_d = 1
  \quad
  \mbox{at the outer sonic point}
  \ .
\end{equation}
The second dark matter boundary condition is the value of the ratio
$\rho_d / \rho_b$
of dark matter to baryonic proper mass densities at the outer sonic point,
which we vary as discussed in \S\ref{DMsec} below.

\section{Results}
\label{ResultsbaryonsDM}

\S{}IV of Paper~1 presented results for black holes
which accrete a single fluid of charged baryons.
The baryons either plunged to the singularity,
or else they dropped through the Cauchy horizon,
but mass inflation did not occur.
This section presents results for black holes
which accrete dark matter in addition to baryons,
with the aim of exploring the phenomenon of mass inflation.

In \S\ref{DMsec},
the dark matter will be assumed to have zero cross-section
for absorption by baryons.
In \S\ref{WMsec},
the dark matter will be given a non-zero cross-section
for absorption by baryons.

As in Paper~1,
geometric units $G = c = \Mc = 1$ are used,
where $\Mc$, equation~(100) of Paper~1,
is the charge-augmented interior mass of the black hole
evaluated at the outer boundary, the outer sonic point.

\subsection{Non-interacting dark matter}
\label{DMsec}

\varsDMfig

Figure~\ref{varsDM}
shows results for a black hole
which accretes both charged, non-conducting baryons,
and neutral, pressureless, non-interacting dark matter.
The dark matter particles here are assumed to be massive,
but the results for massless dark matter particles are quite similar.
With these assumptions,
there are three free parameters set at the outer boundary,
the outer sonic point.
The accretion rate $\eta_s$ is set equal to $0.1$,
the same as adopted in the models of Paper~1,
and the charge-to-mass ratio $Q / \Mc$ is set equal to $0.8$,
the same as adopted in the charged models of Paper~1.
The third parameter, a new parameter,
is the ratio $\rho_d / \rho_b$ of dark matter to baryonic proper mass densities,
which we set equal to $0.1$ at the outer sonic point
\begin{equation}
\label{rhodbsonic}
  {\rho_d \over \rho_b} = 0.1
  \ .
\end{equation}

In an astronomically realistic black hole,
the density of accreted dark matter is expected to be only a small
fraction of the density of accreted baryons,
because whereas baryons can dissipitate energy and angular momentum,
which allows them to funnel on to a black hole,
non- or weakly-interacting dark matter cannot so dissipate.
However, as with the other two free parameters $\eta_s$ and $Q / \Mc$,
we deliberately choose a large value of $\rho_d / \rho_b$
to make it easier to discern its effects
(and to avoid the risk of numerical problems).
At fixed $\eta_s = 0.1$ and $Q / \Mc = 0.8$,
the dark matter to baryonic density ratio is limited to
$\rho_d / \rho_b \lesssim 0.4362$
(see Figure~\ref{mr}),
otherwise there is too much neutral dark matter diluting the baryons,
and the desired charge-to-mass $Q / \Mc = 0.8$ cannot be achieved.

As expected,
Figure~\ref{varsDM}
shows that mass inflation occurs just above the inner horizon.
During mass inflation,
the interior mass $M$ increases by many orders of magnitude
over a modest range of radii,
$r \approx 0.372$--$0.1$.
As the Figure shows,
mass inflation in due course ceases,
for reasons discussed in \S\ref{stop} below.
Before mass inflation sets in,
the solution with dark matter resembles the solution without dark matter,
Figure~4 of Paper~1.
As discussed in \S{}IV\,B of Paper~1,
the geometry of the solution without dark matter in turn resembles,
outside the inner horizon,
that of a vacuum charged black hole,
the Reissner-Nordstr\"om geometry.

After mass inflation has stalled,
the outgoing baryonic and ingoing dark matter fluids
collapse to a spacelike singularity at zero radius.
The Penrose diagram of the black hole is the same as that
shown in Figure~3 of Paper~1.

Figure~\ref{varsDM}
shows that the end of mass inflation coincides roughly with
the homothetic scalar $H$ reaching a minimum in absolute value.
The homothetic scalar $H = ( 1 - V^2 ) /\eta$
is zero at the inner horizon, where $V = -1$,
and can be interpreted as offering
a gauge-invariant measure of how close to the inner horizon
the mass-inflating fluid has reached.
%In \S\ref{stop}
%we will examine the connection
%between mass inflation and the homothetic scalar $H$.

%Figure~\ref{varsDM}
%shows that at much smaller radii $r$,
%the homothetic scalar $H$ eventually rises back above unity in absolute value,
%an indication that the infalling baryons and dark matter are no longer pressing
%the inner horizon.

\mrfig

During mass inflation,
the interior mass $M$ increases approximately exponentially
while the radius $r$ decreases only modestly:
\begin{equation}
  M \sim \exp \left[ - ( \ln r ) / l \right]
\end{equation}
where
%$\Delta \ln r$ is the change in logarithmic radius, and
$l$ is a dimensionless exponential scale length.
The inflationary scale length $l$ is approximately constant during
the main part of mass inflation,
and can be characterized quantitatively by the
reciprocal of the maximum logarithmic derivative of $2 M / r - 1$,
equations~(\ref{rx}) and (\ref{dlnM}), during inflation
\begin{equation}
\label{l}
  {1 \over l}
  \equiv
  \max \left[ - \, {\dd \ln( 2 M / r - 1 ) \over \dd \ln r} \right]
  \ .
\end{equation}
Figure~\ref{mr}
shows the inflationary scale length $l$ defined by equation~(\ref{l})
as a function of the ratio
$\rho_d / \rho_b$
of dark matter to baryonic proper mass density at the outer sonic point,
for black holes with the same accretion rate $\eta_s = 0.1$
and charge-to-mass $Q / \Mc = 0.8$
as that illustrated in Figure~\ref{varsDM}.
Shorter scale lengths $l$ signify more extreme mass inflation.
The scale length increases approximately linearly at small ratios,
$l \approx 0.032 \, \rho_d / \rho_b$,
goes through a maximum
at $\rho_d / \rho_b \approx 0.27$,
and then decreases.
As $l$ declines,
the curve passes through a mild maximum in $\rho_d / \rho_b$,
at $\rho_d / \rho_b \approx 0.4362$,
and terminates at
$\rho_d / \rho_b \approx 0.433$,
constrained by $\gamma^2 - \beta^2 > 0$ at the outer sonic point.

Since the simultaneous presence of outgoing (baryonic) and ingoing (dark matter)
fluids is essential to mass inflation,
one might have thought that mass inflation would be most extreme
when the densities of baryonic and dark matter were comparable.
Figure~\ref{mr}
shows that the opposite is true:
mass inflation is most extreme
(the inflationary scale length $l$ is smallest)
when either the ingoing dark matter stream is reduced to a trace
(small $\rho_d / \rho_b$),
or the outgoing baryonic stream is reduced to a trace
($\rho_d / \rho_b \approx 0.433$).
In the latter case,
the baryonic density decreases by many orders of magnitude
inside the black hole, so that at the onset of mass inflation
the proper density of baryons is indeed only a trace
compared to the proper density of dark matter.

When mass inflation is more extreme in the sense that
the inflationary scale length $l$ is small,
it is also more extreme in the sense that the mass $M$
exponentiates to a larger value before mass inflation ends.

The conundrum that mass inflation is most extreme when
one of the ingoing or outgoing streams is reduced to a trace
is considered in \S\ref{less} below.

The exponential increase of the interior mass $M$
is paralleled by an exponential increase in (the absolute value of) the
streaming 4-velocity $\utet_d^r$ of the dark matter through the baryons,
and an exponential decrease in (the absolute value of) the homothetic scalar $H$
\begin{equation}
  \utet_d^r \sim \exp \left[ - ( \ln r ) / l \right]
  \ , \quad
  H \sim \exp \left[ ( \ln r ) / l \right]
  \ .
\end{equation}
Figure~\ref{mr} is practically unchanged if the inflationary scale length $l$
is defined either by the exponential scale length of the 4-velocity $\utet_d^r$
[see equation~(\ref{dlnu})]
\begin{equation}
  {1 \over l}
  \equiv
  \max \left[ - \, {\dd \ln( \utet_d^r ) \over \dd \ln r} \right]
\end{equation}
or by the exponential scale length of the homothetic scalar $H$
[see eqation~(\ref{dlnH})]
\begin{equation}
  {1 \over l}
  \equiv
  \max \left[ {\dd \ln( H ) \over \dd \ln r} \right]
\end{equation}
in place of equation~(\ref{l}).

\subsection{Interacting dark matter}
\label{WMsec}

A feature of mass inflation is that ingoing and outgoing fluids
stream through each other at ever closer to the speed of light
(or become ever more blueshifted, for massless streams).
This raises the physical question of what happens if there is a finite
cross-section for interaction between the ingoing and outgoing fluids
at sufficiently high collision energies.

We have carried out a number of numerical experiments
in which we have adjusted both the size and dependence on collision energy
of the dimensionless rate coefficient $\kappa_d$
in the absorption rate $\sigma_d$, equation~(\ref{sigmad}),
of dark matter by baryons.
In all cases we find that,
as long as the rate $\sigma_d$ is finite (not infinite) at all energies,
then the results are similar to those found in \S\ref{DMsec}:
mass inflation occurs, then comes to an end,
whereupon the fluids plunge to the central singularity.

As the absorption rate coefficient $\kappa_d$ is varied,
the degree of mass inflation changes,
in a manner consistent with what was found in \S\ref{DMsec}.
For example,
if the parameters are set to those of Figure~\ref{varsDM},
namely
$\eta_s = 0.1$,
$Q / \Mc = 0.8$,
and
$\rho_d / \rho_b = 0.1$,
then mass inflation becomes more extreme
as the absorption rate coefficient $\kappa_d$ is increased,
with the inflationary scale length $l$ shortening,
and the mass $M$ exponentiating to a higher value before inflation ends.
Increasing the absorption rate reduces the density of dark matter,
which has a similar effect to sliding $\rho_d / \rho_b$
in Figure~\ref{mr} to values less than $0.1$,
thereby reducing the inflationary scale length $l$.

Is it possible to adjust the interaction between dark matter and baryons
so that the baryons drop through the Cauchy horizon, as in Paper~1?
The only way to achieve this is to let
the absorption rate become infinite at a finite collision energy.
If the absorption rate is infinite,
then all the dark matter is absorbed and only outgoing baryons remain.
As soon as all the dark matter is gone,
then the baryons can drop through the Cauchy horizon.
This is consistent with the fact,
proven in \S\ref{nodrop} below,
that as long as ingoing and outgoing fluids are simultaneously present,
then it is impossible for either fluid
to drop through the inner horizon.

\varsWMfig

%\shkWMfig

Figure~\ref{varsWM}
shows results for a black hole accreting baryons and dark matter
with a dimensionless absorption rate coefficient $\kappa_d$
that becomes numerically infinite
when the 4-velocity $\utet_d^m$ of the dark matter through the baryons is large
\begin{equation}
\label{kappadWM}
  \kappa_d = 10^{-20} ( \utet_d^t )^2
  \ .
\end{equation}
The rate coefficent $\kappa_d$ given by equation~(\ref{kappadWM})
is of course analytically finite at all collision energies,
but numerically it is large enough at high collision energy
($\utet_d^t \gg 10^{10}$)
that the dimensionless dark matter density $z_d \equiv 4 \pi r^2 \rho_d$
falls below about $10^{-300}$,
at which point $z_d$ underflows and is set to zero by the numerics.
Aside from the fact that $\kappa_d$
%given by equation~(\ref{kappadWM})
becomes large at large $\utet_d^t$,
there is nothing magic about
the constant of proportionality $10^{-20}$
or the exponent $2$ in equation~(\ref{kappadWM}).
The values are chosen simply to yield an interesting amount of mass inflation,
as shown in Figure~\ref{varsWM},
before the outgoing baryons drop through the Cauchy horizon.

As Figure~\ref{varsWM} illustrates,
the structure of the black hole accreting dark matter with
an effectively infinite high-energy absorption rate
is similar to that of non-interacting dark matter, Figure~\ref{varsDM},
up to the point where the dark matter is completely absorbed.
At that point,
the outgoing baryons drop through the Cauchy horizon,
similar to the situation illustrated in Figure~4 of Paper~1,
where there was no dark matter present.

The similarity solution for the infinitely-interacting dark matter model
of Figure~\ref{varsWM}
does not continue consistently to zero radius inside the Cauchy horizon,
but rather terminates at an irregular sonic point at finite radius.
This is the same phenomenon as happened in Paper~1
whenever the baryons dropped through the Cauchy horizon.
As in Paper~1,
the similarity solution of Figure~\ref{varsWM}
terminates whether or not a shock is introduced inside the Cauchy horizon.
The situation is similar to that illustrated in Figure~5 of Paper~1,
and the reader is referred to
\S\S{}IV\,B and IV\,E of Paper~1
for further discussion of this issue.

The Penrose diagram of the black hole of Figure~\ref{varsWM}
is the same as that shown in Figure~6 of Paper~1.

\section{Why mass inflation happens}
\label{Inflation}

This section discusses the physical question of why mass inflation occurs.
\S\ref{nodrop}
shows that,
as long as ingoing and outgoing streams are simultaneously present,
they cannot drop through an inner horizon.
\S\ref{drive}
shows how the gravitational force drives the counter-streaming
of ingoing and outgoing fluids that produces mass inflation.
\S\ref{less}
answers the question of why reducing
one of the ingoing dark matter or outgoing baryonic streams to a trace
relative to the other stream makes mass inflation more extreme,
not the other way round as one might naively have expected.
\S\ref{stop}
considers why mass inflation comes to an end,
as found empirically in \S\ref{DMsec}.
\S\ref{nonull}
discusses the conditions for a null singularity
to form on the Cauchy horizon,
something that does not happen in the similarity solutions.

\subsection{Ingoing and outgoing streams cannot drop through an inner horizon}
\label{nodrop}

The simultaneous presence of ingoing and outgoing fluids
in the vicinity of the inner horizon is,
as first pointed out by Poisson \& Israel (1990) \cite{PI90},
the essential ingredient for mass inflation to occur.
In this subsection we show that,
in the context of the similarity solutions considered in this paper,
as long as ingoing and outgoing fluids are simultaneously present,
then it is impossible for the fluids to drop through the inner horizon,
because to do so would require that the ingoing and outgoing fluids
stream through each other faster than the speed of light,
which is impossible.

Below we will refer to the outgoing fluid as baryons,
and the ingoing fluid as dark matter,
but the argument applies generally to any combination
of ingoing or outgoing fluids.

As described in \S{}III\,C of Paper~1,
in similarity solutions
a frame of reference is ingoing or outgoing
depending on whether the time component $\xi^t$
of the homothetic 4-vector is positive or negative in that frame.
The components of the homothetic vector in the dark matter and baryonic frames
are related by a Lorentz transformation,
equation~(\ref{xid}).
If the proper velocity of the dark matter relative to the baryons
is denoted $\vd \equiv \utet_d^r / \utet_d^t$
%(do not confuse the latin vee $v$, denoting velocity in the tetrad frame,
%with the greek upsilon $\upsilon$, denoting 4-velocity in the coordinate frame)
then the time component $\xi_d^t$ of the homothetic vector
in the dark matter frame is related to the time component $\xi^t$
in the baryonic frame by,
equation~(\ref{xid}),
\begin{equation}
  \xi_d^t = \xi^t \utet_d^t \left( 1 - V \vd \right)
  \ .
\end{equation}
Since $\utet_d^t$ is always positive,
it follows that the dark matter will have the opposite in/out sign from baryons
(viz.\ ingoing, $\xi_d^t > 0$, if the baryons are outgoing, $\xi^t < 0$)
if and only if
\begin{equation}
\label{vv}
  V \vd > 1
\end{equation}
that is, if $\vd > 1/V$ for positive $V$,
or if $\vd < 1/V$ for negative $V$.
We remind the reader that $V$ is the proper velocity of the similarity frame
relative to the baryonic frame,
and that the absolute value of the velocity is equal to one, $|V| = 1$,
at horizons.

Since the velocity $\vd$ of the dark matter relative to the baryonic frame
is necessarily less than or equal to the speed of light, $|\vd| \leq 1$,
it follows from equation~(\ref{vv}) that
dark matter can have the opposite in/out sign from the baryonic frame
only in superluminal regions of the geometry, where $|V| > 1$.
Ingoing and outgoing fluids cannot coexist
in the same contiguous subluminal region, where $|V| < 1$,
since to do so they would have to move faster than light
relative to each other.
If ingoing and outgoing fluids fall
through the inner horizon,
then they must necessarily pass
into separate ingoing and outgoing subluminal regions.

Approaching the inner horizon,
where $|V| \rightarrow 1$,
ingoing and outgoing frames
must necessarily approach the speed of light,
$|\vd| \rightarrow 1$,
relative to each other, according to equation~(\ref{vv}).
At the inner horizon, $|V| = 1$,
ingoing and outgoing objects
must necessarily
stream through each other at the speed of light, $|\vd| = 1$,
which is problematic if the objects have finite rest mass.
Indeed it is problematic even if the objects have zero rest mass,
because, at the inner horizon,
a light ray which has finite energy in an ingoing frame
must appear infinitely blueshifted in an outgoing frame.

The above two paragraphs have demonstrated the claimed assertion,
that as long as ingoing and outgoing fluids are simultaneously present,
then it is impossible for the fluids to drop through the inner horizon,
because to do so they would have to exceed the speed of light.
A corollary of the argument is that as soon as one of the streams is exhausted,
then the other stream can drop through the inner horizon,
an ingoing horizon if only ingoing fluid remains,
or an outgoing horizon, the Cauchy horizon, if only outgoing fluid remains.

\subsection{Gravity drives mass inflation}
\label{drive}

The previous subsection, \S\ref{nodrop},
showed that for ingoing and outgoing fluids
to approach an inner horizon, they must stream ever faster through each other.
But what drives such counter-streaming?

The thing that drives ingoing and outgoing fluids to stream
ever faster through each other is the inward gravitational force.
The trick is that `inward',
meaning in the direction of smaller radius $r$,
means opposite directions for the ingoing and outgoing fluids.
Ingoing and outgoing fluids are both accelerated inwards,
but nevertheless they are accelerated in opposite directions.

In \S{}IV\,B of Paper 1
we attached a gyroscope to an infalling observer,
and, having initialized the gyroscope so that it points towards the black hole,
we defined the direction in which the gyroscope points
as the immutable direction towards the black hole.
In the locally inertial (tetrad) frame of the infalling observer,
the direction towards the black hole
is not necessarily the direction of smaller circumferential radius $r$.
Rather,
the direction of smaller circumferential radius $r$
is determined by the sign of the vierbein coefficient
$\gamma \equiv \partial_r r$,
equation~(12) of Paper~1.
A positive $\gamma$ means that
the gyroscope points in the direction of
smaller proper circumferential radii $r$.
Conversely,
a negative $\gamma$ means that
the gyroscope points in the direction of larger proper circumferential radii.
Zero $\gamma$ means that
the circumferential radius is an extremum.

The reader who is not yet convinced that accelerating inwards,
i.e.\ making the radial coordinate velocity
$\beta \equiv \partial_t r$
more negative,
can mean accelerating in two opposite directions
is invited to consider what happens to the radial 4-gradient
$(\beta, \gamma) \equiv (\partial_t r, \partial_r r)$
under accelerations,
i.e.\ under Lorentz boosts, equation~(\ref{bgud}),
in the case at hand,
where the radial 4-gradient is timelike, $\beta^2 - \gamma^2 > 0$,
and $\beta$ is negative.

The sign of the vierbein coefficient $\gamma$
is not in one-to-one correspondence with
whether the fluid is ingoing or outgoing
(cf.\ Table~II of Paper~1),
but the two do correspond
just outside the inner horizon:
$\gamma$ is positive if the fluid is ingoing,
negative if the fluid is outgoing.

%Gravity drives counter-streaming of ingoing and outgoing fluids
%even in the case of a vacuum black hole, the Reissner-Nordstr\"om (RN) geometry.
%In the RN geometry,
%ingoing and outgoing test particles see each other infinitely blueshifted
%as they drop through their respective ingoing and outgoing inner horizons.
%Thus the seed of mass inflation---a
%streaming velocity that tends to the speed of light at the inner
%horizon---is already built into the RN geometry.

%As the RN counter-example shows,
%a large streaming velocity is not by itself sufficient to drive mass inflation.
What drives mass inflation is a feedback loop
in which the increasing radial pressure of the counter-streaming fluids
amplifies the gravitational force, which accelerates the counter-streaming,
which in turn increases the radial pressure.
To see how this works,
consider the gravitational force equation~(\ref{accelerationd})
which governs the acceleration $\partial_t \beta$
of the coordinate radial velocity $\beta \equiv \partial_t r$.
Equation~(\ref{accelerationd}) is nominally for the baryonic tetrad frame,
but essentially the same equation remains valid in the dark matter tetrad frame,
if all quantities in the equation are reinterpreted as
relative to the dark matter frame.

The gravitational force equation~(\ref{accelerationd})
expresses the radial acceleration $\partial_t \beta$ as a sum of three terms:
the familiar attractive Newtonian force $- M / r^2$,
an additional general relativistic gravitational force $- 4 \pi r T^{rr}$
whose source is the radial pressure $T^{rr}$,
and a force $\gamma g$ which comes from the acceleration generated by
pressure balance
(which includes the Lorentz force),
and whose presence expresses the Principle of Equivalence.
Numerically (\S\ref{DMsec}),
the dominant term during mass inflation,
for both baryons and dark matter,
is the inward gravitational force term $- 4 \pi r T^{rr}$.
%The Newtonian contribution $- M / r^2$ to the gravitational force
%plays a role in helping accelerate the counter-streaming fluids
%to the point where mass inflation can begin,
%but by itself would not be sufficient to cause mass inflation,
%as the RN counter-example shows.
%Rather,
%it is
It is this general relativistic force $- 4 \pi r T^{rr}$,
which increases as the square of the streaming radial 4-velocity $\utet_d^r$,
that drives mass inflation.
By contrast, the Newtonian contribution $- M / r^2$, to the gravitational force
increases only linearly with the streaming radial 4-velocity $\utet_d^r$.
In the baryons the inward gravitational force
is partially opposed by the $\gamma g$ force
from a strong pressure gradient generated
as a back reaction to the gravitational force,
but still the primary gravitational force wins.
The dark matter is pressureless,
so in that case the $\gamma g$ force is zero.

%One should not think that
%ingoing and outgoing fluids pause near the inner horizon while
%they undergo mass inflation.
%On the contrary,
%both ingoing and outgoing fluids accelerate strongly inwards.
%For both baryons and dark matter,
%all of mass inflation and after takes place in flash of proper time.

It is curious that the gravitational force term
$- 4 \pi r T^{rr}$
is the key player in two seemingly opposite roles.
On the one hand,
in a vacuum charged black hole this gravitational force term is repulsive,
thanks to the negative radial pressure of the electric field.
It is this gravitational repulsion that causes a vacuum black hole
to contain an inner horizon,
where ingoing and outgoing frames accelerate to the speed of light
relative to each other.
Without the gravitational repulsion,
ingoing and outgoing fluids would not be inclined to accelerate
through each other to the speed of light,
and mass inflation would not begin.
On the other hand,
it is this same gravitational force term,
with an exponentially growing positive pressure rather
than a passive negative pressure,
that provides the feedback loop that drives mass inflation.

%It is ironic that the gravitational force term
%$- 4 \pi r T^{rr}$
%that makes vacuum charged black holes repulsive in their cores,
%thanks to the negative radial pressure of the electric field,
%is the same gravitational force term that grows hugely attractive
%during mass inflation.

The electric force behaves differently from the gravitational force.
Irrespective of the sign of $\gamma$,
electrically charged particles can consistently interpret the electric force
as being caused
either by a positive charge $Q$ located in the direction of the black hole,
or by a negative charge $-Q$ located in the direction away from the black hole.
Either way,
a positively charged particle is always repelled in the direction away
from the positively charged black hole.

\subsection{Why is less more?}
\label{less}

In \S\ref{DMsec}
it was found that reducing
one of the ingoing dark matter or outgoing baryonic streams to a trace amount
relative to the other stream
actually resulted in more extreme mass inflation,
in the sense that the inflationary scale length $l$,
equation~(\ref{l}), became shorter,
and the mass $M$ exponentiated to a larger value,
before mass inflation came to an end.
An analogous result was found in \S\ref{WMsec},
where for example increasing the absorption rate of dark matter by baryons
so that the ingoing dark matter stream was reduced to a trace
again led to more extreme mass inflation.
Since the simultaneous presence of both ingoing and outgoing streams
is a prerequisite for mass inflation to occur,
one might have thought that comparable amounts of
ingoing and outgoing fluid would produce more inflation.
But the opposite is true:
less of one stream relative to the other produces more inflation,
with comparable amounts of ingoing and outgoing fluid producing
the least inflation.
Why?

For definiteness,
consider the (more realistic) case where the dark matter density
is only a small fraction of the baryonic density
(an analogous argument applies in the opposite case where the baryonic density
is only a small fraction of the dark matter density).
Mass inflation begins at around the time that the general relativistic
gravitational force $- 4 \pi r T^{rr}$ becomes dominant
in the acceleration equation~(\ref{accelerationd}),
which happens when the radial streaming 4-velocity $\utet_d^r$
has become large enough in absolute value.
The smaller the dark matter density,
the larger the 4-velocity $\utet_d^r$ must become in order
for $- 4 \pi r T^{rr}$ to take over as the dominant gravitational force.
A larger 4-velocity $\utet_d^r$ requires that the ingoing and outgoing fluids
approach closer to the inner horizon before mass inflation begins.
As the fluids approach the inner horizon,
the characteristic scale length $l$ over which the 4-velocity $\utet_d^r$
increases becomes shorter and shorter.
When in due course mass inflation sets in,
it is this characteristic scale length that
determines the scale length over which $M$ and $\utet_d^r$
then exponentiate together.
A smaller dark matter density
thus implies a smaller inflationary length scale $l$,
hence more extreme inflation.

\subsection{Why does mass inflation end?}
\label{stop}

In \S\ref{DMsec}
it was found that mass inflation eventually ceased,
whereafter the ingoing and outgoing fluids
plunged to a spacelike singularity at zero radius.
Why does mass inflation end?

What happens is that the streaming 4-velocity $\utet_d^r$
ceases its exponential growth,
and indeed starts shrinking instead of growing.
The dominant term in the differential equation~(\ref{dlnu})
governing the evolution of the 4-velocity $\utet_d^r$
is the term proportional to $T_\xi^{tr}$,
which can be interpreted as the direct gravitational force term.
The other term,
proportional to the acceleration $y \equiv g r$
generated as a result of pressure balance,
slightly counteracts the direct $T_\xi^{tr}$ term, but not much.
Now $T_\xi^{tr}$
is the momentum density in the no-going $\xi^t = 0$ frame of reference,
at the border between ingoing frames (positive $\xi^t$)
and outgoing frames (negative $\xi^t$).
The momentum density $T_\xi^{tr}$, equation~(\ref{Txi}),
is a sum of two opposing terms,
one from the outgoing baryons,
the other from the ingoing dark matter.
At the outset of inflation,
the contribution to $T_\xi^{tr}$ from the baryons
exceeds that from the dark matter.
As mass inflation continues,
the relative contribution from dark matter gradually becomes more important.
This is because as time goes by,
the dark matter streaming through the baryons
is accreted later and later in the evolution of the black hole,
and in the similarity solutions the mass of accreted dark matter
increases linearly with time.
Eventually,
the contribution to $T_\xi^{tr}$ from the ingoing dark matter
exceeds that from the outgoing baryons,
and $T_\xi^{tr}$ switches sign.
Equivalently,
the center-of-mass frame, where $T^{tr} = 0$,
switches from outgoing to ingoing.
At this point, or actually just before this point
thanks to the $y \equiv g r$ term,
the streaming 4-velocity $\utet_d^r$ starts shrinking instead of growing,
and mass inflation has come to an end.

We must admit that we find it physically somewhat mysterious
that the gravitational force can operate in different directions
on the ingoing and outgoing fluids, as argued in \S\ref{drive},
and yet the proper streaming 4-velocity $\utet_d^r$ can nevertheless shrink.
We can only assume that this mystery can be attributed to
the difference between coordinate velocities and proper velocities.
Whatever the case, the mathematics governing $\utet_d^r$, equation~(\ref{dlnu}),
is clear enough.

The behavior of the homothetic scalar $H$ during inflation
is closely related to that of the streaming 4-velocity $\utet_d^r$.
The equation~(\ref{dlnH}) governing the evolution of $H$
contains, like equation~(\ref{dlnu}) governing $\utet_d^r$,
two terms:
a term proportional to the momentum density $T_\xi^{tr}$ in the no-going frame,
and another term [the one proportional to $(\gamma \xi^t + \beta \xi^r) / H$]
which during inflation is subdominant and acts in mild opposition
to the principal $T_\xi^{tr}$ term.
As with $\utet_d^r$,
the end of inflation is signaled by $T_\xi^{tr}$ changing sign,
and accordingly $H$,
which decreased exponentially in absolute value during inflation,
starts rising back up again.
This is precisely what was found empirically in \S\ref{DMsec},
where inflation stalled at about the time that the homothetic scalar $H$
passed through a minimum in absolute value.

In summary,
inflation comes to an end when the center-of-mass frame
of the counter-streaming fluids switches from outgoing to ingoing.
This condition is true
even in the case where the baryonic density is a trace compared
to the dark matter when mass inflation begins.
This case corresponds to the small inflationary scale lengths $l$
attained at the right edge of Figure~\ref{mr},
where the accreted dark matter to baryonic density
$\rho_d / \rho_b$
is near maximal.
As mentioned in the commentary to Figure~\ref{mr},
even though the dark matter and baryonic densities are comparable
at the sonic point boundary in this case,
$\rho_d / \rho_b \approx 0.433$ for the parameters of Figure~\ref{mr},
the baryonic density decreases by many orders of magnitude
inside the black hole,
so that indeed the baryonic density is driven to a trace compared
to the dark matter density by the time mass inflation begins.
But while the baryonic density diminishes,
the components $\xi^m$ of the homothetic vector in the baryonic frame grow,
with the net result that the baryonic contribution
$(1 {+} w) z \xi^t \xi^r$
to the momentum density $T_\xi^{tr}$
exceeds the dark matter contribution
$z_d \xi_d^t \xi_d^r$
at the onset of inflation, even though $z \ll z_d$.
As mass inflation continues,
the dark matter contribution to the momentum density $T_\xi^{tr}$
grows relatively more important, as in the usual case.
Eventually,
the contribution to $T_\xi^{tr}$ from the ingoing dark matter
exceeds that from the outgoing baryons,
$T_\xi^{tr}$ switches sign,
%or equivalently the center-of-mass frame switches from outgoing to ingoing,
and mass inflation comes to an end.

\subsection{Null singularity on the Cauchy horizon?}
\label{nonull}

Many previous papers have found that the collapse of a massless scalar
field into a charged black hole produces
not only a strong spacelike singularity at zero radius
but also a weak null singularity at finite radius along the Cauchy horizon
\cite{Ori91,BDIM94,BS95,Burko97,BO98,HP98b,HP99,Ori99,HKN05}.
Indeed,
\cite{HKN05}
find that two distinct null singularities may form,
one ingoing and one outgoing.
However,
Burko \cite{Burko02a,Burko03}
finds numerically that a null singularity forms only if the scalar field
set up outside the horizon falls off sufficiently rapidly,
the required degree of rapidity depending on the parameters of the problem,
such as the charge-to-mass ratio of the black hole.
If too much scalar field continues to be accreted,
then no null singularity forms,
and the field collapses to a central singularity.
In the earliest numerical simulation,
Gnedin \& Gnedin (1993) \cite{GG93}
found only a spacelike central singularity, no null singularity,
and it seems likely that the initial conditions for their scalar field
exceeded the Burko bound
(as opposed to there being some flaw \cite{BS95,Burko97}
in the \cite{GG93} method).

A null singularity does not form in the similarity solutions,
and it may be presumed that this is
because the assumption of self-similarity precludes it.
As described in \S\ref{stop},
mass inflation ceases once sufficient dark matter has been
accreted that the center of mass frame near the inner horizon
becomes ingoing rather than outgoing.
At this point,
the outgoing baryons do not persist at finite radius,
but rather collapse to a spacelike singularity at zero radius.
The self-similar hypothesis requires that
the mass of accreted dark matter increase linearly with time
into the indefinite future,
so that outgoing baryonic fluid must inevitably encounter, sooner or later,
enough ingoing dark matter to bear it down to the central singularity.

The methods of the present paper,
which is restricted to self-similar solutions,
are insufficiently powerful
to answer definitely the question of what circumstances
lead to a null singularity on the Cauchy horizon
in the general, non-self-similar case.
However,
the results do suggest that two criteria may be key.
We state these two criteria below in the form of conjectures,
couched in somewhat loose language.

The first conjectured key criterion
for the formation of a null singularity on the Cauchy horizon
is that the amount of ingoing fluid accreted by the black hole should
be finite and `sufficiently small'.
This comes from the empirical finding
that the counter-streaming ingoing and outgoing fluids collapse
to a singularity once sufficient ingoing fluid has been accreted.
Roughly, collapse happens when the mass of accreted ingoing fluid
is comparable to the mass of accreted outgoing fluid,
although we cannot be sure that this approximate criterion is true in general.

To avoid confusion,
we should comment that by ingoing and outgoing accreted fluid,
we mean fluid whose properties are such that it becomes ingoing or outgoing
near the inner horizon.
The term accretion is also intended in a loose sense.
For example, ingoing fluid could possibly emit outgoing fluid, or vice versa,
so the source of ingoing and outgoing fluids
may not necessarily be accretion.

The second conjectured key criterion is that
the black hole should accrete ingoing fluid into the indefinite future.
This comes from the idea that if the accretion of ingoing fluid is cut short,
then the outgoing fluid will run through all the available ingoing fluid,
and will then promptly drop through the Cauchy horizon.

It should also be commented that the definition of ingoing versus outgoing
adopted in this paper and in Paper~1,
that a frame is ingoing or outgoing according to whether the time
component $\xi^t$ of the homothetic 4-vector is positive or negative,
works only for self-similar solutions,
since the homothetic vector exists only in self-similar solutions.
An alternative definition
that would work in a general spherically symmetric metric
would be to define a frame as ingoing or outgoing according to the sign
of the vierbein coefficient $\gamma$.
This definition agrees with our adopted definition
in the important mass-inflationary region near the inner horizon.
Indeed, it has been seen in \S\ref{drive}
that the sign of $\gamma$
is physically at the heart of mass inflation,
because it is is the sign of $\gamma$
that determines which direction is `inward'
(meaning the direction of smaller circumferential radius $r$),
and therefore in which direction the gravitational force operates,
towards the black hole for ingoing fluid (positive $\gamma$),
and away from the black hole for outgoing fluid (negative $\gamma$).

\appearfig

\section{Appearance of the black hole}
\label{Appearance}

\S{}V of Paper~1 considered the question:
What does it actually look like if you fall inside one of the black holes
described in that paper?
This section addresses the same question
for the black holes considered in the present paper.

Figure~\ref{appear} shows,
for the two models illustrated in Figures~\ref{varsDM} and \ref{varsWM},
the angular size $\chi_\textrm{ph}$
and blueshift of photons from the edge of the black hole,
as observed either in the baryonic rest frame
or in the radially free-falling dark matter rest frame.
The two points of view are related by a radial Lorentz boost.
The observed angular size $\chi_\textrm{ph}$ of the black hole
(the subscript ph signifying photons from the photon sphere equivalent)
is given by equation~(73) of Paper~1,
and the observed blueshift of photons at the edge of the black hole
is given by equation~(74) of Paper~1.
For the free-fall dark matter frame,
equations~(73) and (74) of Paper~1
apply with the baryonic homothetic vector $\xi^m$
replaced by its dark matter counterpart $\xi_d^m$,
equation~(\ref{xid}).
The horizontal axis on Figure~\ref{appear} is the radius as measured
in the corresponding frame,
$r$ for the baryons, $r_d$ for the dark matter.

Figure~\ref{appear}
shows that,
down to the point where mass inflation begins,
the appearance of black holes
accreting dark matter in addition to charged baryons
is similar to the appearance of black holes accreting only charged baryons,
middle panel of Figure~12 of Paper~1.
As discussed in \S{}V of Paper~1, this appearance
is in turn similar to that of the corresponding vacuum black hole,
the Reissner-Nordstr\"om solution.

From the point of view of an outgoing observer,
such as one in the baryonic rest frame,
the black hole
(that is, any of the black holes considered in this paper)
increases in angular size until it covers almost the entire sky.
The view of the outside universe correspondingly shrinks to
a small, intensely bright, blueshifted point above the observer.
As long as mass inflation continues,
the point gets smaller, brighter, and more blueshifted.

From the point of view of an ingoing observer on the other hand,
such as one in the free-fall dark matter rest frame,
the black hole
(again meaning any of the black holes considered in this paper)
first increases in angular size,
but then shrinks as the observer approaches the inner horizon.
In contrast to the outgoing observer who sees the outside universe concentrate
to a small point,
the ingoing observer sees the outside universe cover almost the whole sky.
To the ingoing observer,
the sky near the edge of the black hole appears blueshifted,
but the sky away from the edge is mostly redshifted.
During mass inflation,
the black hole continues to shrink,
and to be surrounded by a concentrating, brightening halo.

In models where the dark matter is non-interacting,
such as the model shown in the left panel of Figure~\ref{appear},
or where the dark matter has a finite (not infinite) cross-section
for absorption by baryons,
mass inflation eventually ceases,
as discussed in \S\S\ref{DMsec} and \ref{stop}.
As mass inflation comes to an end,
an outgoing observer, such as one in the baryonic frame,
sees the view of the outside universe start to re-expand,
%(initially barely perceptibly),
and to become less bright and less blueshifted.
As the outgoing baryons plunge to the singularity at zero radius,
their view of the outside universe expands to a radius of $90^\circ$.
The $90^\circ$ view near the singularity is similar
to that seen by an observer who falls into a Schwarzschild black hole,
and can be attributed to the same enormous tidal force
that stretches the infaller radially and crushes them horizontally.
As in the Schwarzschild solution,
the blueshift at the edge of the black hole
tends to infinity as the outgoing baryonic observer approaches the singularity,
but the amount of time that the observer sees pass by in the outside universe,
the integral of blueshift over proper time, is finite.

As remarked above,
an ingoing observer sees a different view,
a tiny black hole surrounded by a bright, blueshifted halo.
The ingoing observer
sees the halo become more concentrated, brighter, and more blueshifted,
not only during mass inflation, but also thereafter,
all the way down to the singularity at zero radius.
Although the blueshift around the black hole tends to infinity
at the singularity,
the ingoing observer sees, like the outgoing observer,
only a finite time pass by in the outside universe.

In models where the dark matter absorption rate is effectively infinite
at high energy,
such as the model shown in the right panel of Figure~\ref{appear},
the outgoing baryons drop though the Cauchy horizon
as soon as the ingoing dark matter is completely absorbed.
If the outgoing baryons could see the outside universe
(which they cannot, because by assumption there is no ingoing
matter or radiation left to see),
then as the outgoing baryons dropped through the Cauchy horizon,
their view of the outside universe would disappear
in an infinitely bright, blueshifted, concentrated flash,
in which the entire future of the outside universe passes by.

The right panel of Figure~\ref{appear}
shows the view seen by observers
in the freely-falling ingoing dark matter frame,
cut short once the dark matter has been completely absorbed.
Cutting the view short seems natural since the model is specifically constructed
so that there is no ingoing fluid---no dark matter---beyond a certain point.
However,
if there were ingoing test particles
(with vanishing energy-momentum tensor),
then their view would resemble that illustrated in the middle
panel of Figure~12 of Paper~1,
in which the ingoing test particles fall to zero radius,
encountering outgoing baryons accreted at ever earlier times,
while the outgoing particles' view of the outside universe
blueshifts to infinity.

As already mentioned,
in most cases observers see only a finite time go by
in the outside universe as they voyage to their doom inside the black hole.
The exception is that
if an outgoing observer drops through the Cauchy horizon,
then, if the outgoing observer could see the outside universe,
the observer would see the entire future of the universe pass by.
Of course,
the outgoing observer can drop through the Cauchy horizon only if
there is no ingoing fluid left,
in which case the observer cannot see the outside universe.
In effect then,
there is no case in which an infalling observer ever sees
an infinite future go by.

\radDMfig

For the black holes illustrated in Figure~\ref{appear},
the amount of time that an infalling observer
sees during their voyage into the black hole is quite modest.
The baryons and dark matter that stream through each other inside the black
hole were accreted at different ages $t$ and $t_d$,
and Figure~\ref{radDM} shows,
for the two models illustrated in Figures~\ref{appear},
the ratio $t_d/t$ of these two ages at each point inside the black hole.
The ratio of ages equals the reciprocal of the ratio of radii,
$t_d/t = r/r_d$,
computed from equations~(\ref{rx}) and (\ref{rdx}).
For the non-interacting model,
the ratio of ages tends to $t_d/t \rightarrow 2.034$ as $r \rightarrow 0$.
The ratio is only slightly different
if the dark matter is massless instead of massive,
$t_d/t \rightarrow 2.007$ as $r \rightarrow 0$
(photons that do not scatter off baryons
can be regarded as a kind of massless dark matter).
The ratio $t_d/t \approx 2$ as $r \rightarrow 0$
means that the baryons see a factor of two into the future:
the baryons see dark matter which was accreted when the black hole was
twice as old as when the black hole accreted the baryons.
Similarly,
the dark matter sees a factor of two into the past:
the dark matter sees baryons which were accreted when the black hole was
half as old as when the black hole accreted the dark matter.

Roughly speaking,
mass inflation ceases and the fluids collapses to a singularity
when comparable masses of outgoing baryons and ingoing dark matter
have been accreted.
Thus the ratio $t_d/t$ of ages would be larger
if the ratio $\rho_d / \rho_b$ of accreted dark matter to baryonic density
were reduced.

Figure~\ref{radDM}
shows that the case of the infinitely-interacting dark matter
is similar to that of the non-interacting dark matter
until the dark matter is completely absorbed,
beyond which there is no dark matter left to see, or to be seen by, the baryons.

At the end of \S{}V of Paper~1
it was remarked that the function $H(X)$,
which plays the essential part in ray-tracing,
equation~(70) of Paper~1,
was reasonably approximated as a cubic or quartic polynomial in $X$.
While this approximation remains satisfactory before mass inflation starts,
it fails completely once mass inflation sets in.
During and after mass inflation, $X$ hardly varies at all,
while $H$, as illustrated in Figure~\ref{varsDM},
varies by many, many orders of magnitude.

\section{Summary}
\label{Summary}

In this the second of two companion papers,
we have investigated self-similar solutions
for spherically symmetric charged black holes
that accrete a pressureless fluid of neutral dark matter
(massive or massless)
in addition to
a relativistic fluid ($p_b / \rho_b = 1/3$) of charged baryons.
The primary aim has been to investigate mass inflation.

%Having set up the equations
%in \S\S\ref{Equations} and \ref{Similaritysolutions},
%we presented numerical results in \S\ref{ResultsbaryonsDM},
%and then discussed the physical causes underlying mass inflation
%in \S\ref{Inflation}.

As first pointed out by \cite{PI90},
the essential ingredient of mass inflation is the simultaneous
presence of ingoing and outgoing fluids near the inner horizon.
In the present paper,
the accreted charged baryonic fluid,
repelled by the charge of the black hole
generated self-consistently by previously accreted charged baryons,
naturally becomes outgoing.
The accreted dark matter,
which is neutral, remains ingoing.
Relativistic counter-streaming between outgoing baryons and ingoing dark matter
then leads to mass inflation near the inner horizon, as expected.

\S\ref{Inflation}
discussed the physical causes underlying mass inflation.
In \S\ref{nodrop}
we showed that,
in the context of the similarity solutions considered in this paper,
as long as ingoing and outgoing fluids are simultaneously present,
then it is impossible for the fluids to drop through the inner horizon,
because in order to do so the fluids would have to stream through each other
faster than light, which is impossible.
A corollary of this argument is that,
if either of the ingoing or outgoing streams is exhausted,
then the other stream can promptly drop through the inner horizon,
an ingoing horizon if only ingoing fluid is present,
or an outgoing (Cauchy) horizon if only outgoing fluid is present.

As argued in \S\ref{drive},
the thing that drives ingoing and outgoing fluids to stream
ever faster through each other during mass inflation
is the inward gravitational force.
The trick is that, in the region near the inner horizon,
`inward', meaning in the direction of smaller circumferential radius $r$,
means opposite directions for the ingoing and outgoing fluids.
For ingoing fluid,
the direction of smaller radius points towards the black hole,
whereas for outgoing fluid,
the direction of smaller radius points away from the black hole.
%Initially, the inward gravitational force is produced by the
%familiar Newtonian force $M / r^2$,
%but as mass inflation gets under way the gravitational force becomes
%dominated by the general relativistic contribution
%sourced by the radial pressure,
%which increases rapidly as the counter-streaming fluids
%accelerate ever faster through each other.

In \S\ref{drive},
we remarked on the curious dual role played by the pressure contribution
to the gravitational force.
On the one hand,
it is the negative radial pressure of the electric field
that produces the gravitational repulsion
that decelerates the inward flow of space into the black hole,
and that therefore causes a vacuum black hole to contain an inner horizon.
Without this negative pressure,
there would be no inner horizon,
and no mass inflation.
On the other hand,
the same pressure contribution to the gravitational force,
with an exponentially growing positive pressure rather
than a passive negative pressure,
provides the feedback loop that drives mass inflation.

Since the simultaneous presence of outgoing (baryonic) and ingoing (dark matter)
fluids is essential to mass inflation,
one might have thought that mass inflation would be strongest
in black holes which accrete comparable amounts of baryonic and dark matter.
In the numerical experiments presented in \S\ref{ResultsbaryonsDM}
we found that, on the contrary,
mass inflation becomes more extreme
as one of the ingoing or outgoing streams is reduced to a trace
relative to the other.
Thus, paradoxically, there is a huge difference between the case of
no dark matter (considered in Paper~1)
and the case of a tiny trace of dark matter.
With no dark matter, the baryons can drop quietly through the Cauchy horizon.
With a trace of ingoing dark matter,
the outgoing baryons cannot drop through the Cauchy horizon,
and instead undergo extravagant mass inflation.

In the similarity solutions considered in the present paper,
mass inflation does not continue to an arbitrarily large value of the
interior mass $M$, but rather comes to an end.
Mass inflation ends at approximately the time that the center-of-mass frame
of the counter-streaming fluids switches from outgoing to ingoing,
whereafter the fluids collapse to a spacelike singularity at zero radius.
It has widely been considered that a generic consequence
of mass inflation is a weak null singularity on the Cauchy horizon,
and this conclusion is undoubtedly valid in the situation
(different from that considered in the present paper)
where the outgoing fluid is a power-law tail
\cite{Price72,HP98a,Dafermos04}
of radiation generated when the black hole first collapses.
More recently Burko \cite{Burko02a,Burko03},
reporting numerical experiments on the collapse of a massless scalar field
into a charged black hole,
found that a null singularity forms only if the amplitude of the scalar
field falls off sufficiently rapidly.
It is
%difficult to assess
not clear
whether Burko's criterion is essentially the
same as that found here, but the results are at least consistent.

A feature of mass inflation is that the streaming velocity
between ingoing and outgoing fluids can reach huge Lorentz gamma factors,
which raises the question of whether it is physically plausible
to allow relativistic streaming at immense energies.
In \S\ref{WMsec}
we explored numerically the consequences of allowing the dark matter
to have a finite cross-section for being absorbed by the baryons.
Consistent with the conclusion that
mass inflation becomes more extreme as the amount of dark matter is reduced
to a trace, we find that increasing the absorption rate
merely makes mass inflation more extreme.
The only caveat to this conclusion is that
if the absorption rate is infinite above some collision energy,
then the ingoing dark matter is absorbed completely,
whereupon the outgoing baryons can drop promptly through the Cauchy horizon.

In \S\ref{Appearance}
we discussed what an observer who falls inside one of the black holes
considered in this paper would see.
Among other things,
we found that in all cases an infalling observer
witnesses only a finite amount of time pass by in the outside universe.
This is true even for an observer who drops through the Cauchy horizon,
because although such an observer would,
if they could see the outside universe,
see the entire future of the universe pass by as they dropped
through the Cauchy horizon,
in fact the observer cannot see the outside universe,
because photons that come unscattered from the outside universe are necessarily
ingoing, and an outgoing observer cannot pass through the Cauchy horizon
as long as there is any trace of ingoing matter or radiation.

\begin{acknowledgements}
This work was supported in part by
NSF award ESI-0337286.
\end{acknowledgements}


\section*{References}

\begin{thebibliography}{}

\bibitem{BDIM94}
A. Bonanno, S. Droz, W. Israel and S. M. Morsink,
``Structure of the spherical black hole interior'',
%gr-qc/9411050
Proc.\ Roy.\ Soc.\ London A\ \textbf{450}, 553--67 (1994).

\bibitem{Brady95}
P. R. Brady,
``Self-similar scalar field collapse: naked singularities and critical behavior'',
%gr-qc/9409035
Phys.\ Rev.\ D\ \textbf{51}, 4168--76 (1995).

\bibitem{BS95}
P. R. Brady and J. D. Smith,
``Black hole singularities: a numerical approach'',
%gr-qc/9506067
Phys.\ Rev.\ Lett.\ \textbf{75}, 1256--9 (1995).

\bibitem{Burko97}
L. M. Burko,
``Structure of the black hole's Cauchy horizon singularity'',
%gr-qc/9710112
Phys.\ Rev.\ Lett.\ \textbf{79}, 4958 (1997).

\bibitem{Burko99}
L. M. Burko,
``Singularity deep inside the spherical charged black hole core'',
Phys.\ Rev.\ D\ \textbf{59}, 024011 (1999).

\bibitem{Burko02a}
L. M. Burko,
``Survival of the black hole's Cauchy horizon under noncompact perturbations'',
Phys.\ Rev.\ D\ \textbf{66}, 024046 (2002).

%\bibitem{Burko02b}
%L. M. Burko,
%``Can a black hole be used as a portal to other universes?'',
%Int.\ J.\ Mod.\ Phys.\ D\ \textbf{11}, 1561 (2002).

\bibitem{Burko03}
L. M. Burko,
``Black-hole singularities: a new critical phenomenon'',
%gr-qc/0209084
Phys.\ Rev.\ Lett.\ \textbf{90}, 121101 (2003);
erratum in
Phys.\ Rev.\ Lett.\ \textbf{90}, 249902 (E).

\bibitem{BO98}
L. M. Burko and A. Ori,
``Analytic study of the null singularity inside spherical charged black holes'',
%gr-qc/9711032
Phys.\ Rev.\ D\ \textbf{57}, R7084 (1998).

\bibitem{CC99}
B. J. Carr and A. A. Coley,
``Self-similarity in general relativity'',
Class.\ Quant.\ Grav.\ \textbf{16}, R31--71 (1999).

\bibitem{Christodoulou86}
D. Christodoulou,
``The problem of a self-gravitating scalar field''
Commun.\ Math.\ Phys.\ \textbf{105}, 337--61 (1986);
``Global existence of generalized solutions of the spherically symmetric Einstein-scalar equations in the large'',
Commun.\ Math.\ Phys.\ \textbf{106}, 587--621 (1986);
``The structure and uniqueness of generalized solutions of the spherically symmetric Einstein-scalar equations'',
Commun.\ Math.\ Phys.\ \textbf{109}, 591--611 (1987);
``A mathematical theory of gravitational collapse'',
Commun.\ Math.\ Phys.\ \textbf{109}, 613--47 (1987).

\bibitem{Dafermos03}
M. Dafermos,
``The interior of charged black holes and the problem of uniqueness in general relativity'',
gr-qc/0307013 (2003);
M. Dafermos and I. Rodnianski,
``A proof of Price's law for the collapse of a self-gravitating scalar field'',
gr-qc/0309115 (2003).

\bibitem{Dafermos04}
M. Dafermos,
``Price's law, mass inflation, and strong cosmic censorship'',
in {\sl Proceedings of the Seventh Hungarian Relativity Workshop},
to appear,
gr-qc/0401121 (2004).

%\bibitem{G05}
%Roberto Giamb\'o,
%``Gravitational collapse of homogeneous scalar fields'',
%gr-qc/0501013 (2005).

\bibitem{GG93}
M. L. Gnedin and N. Y. Gnedin,
``Destruction of the Cauchy horizon in the Reissner-Nordstrom black hole'',
Class.\ Quant.\ Grav.\ \textbf{10}, 1083--1102 (1993).

\bibitem{GP87}
D. S. Goldwirth and T. Piran,
``Gravitational collapse of massless scalar field and cosmic censorship'',
Phys.\ Rev.\ D\ \textbf{36}, 3575--81 (1987).

\bibitem{Paper1}
A. J. S. Hamilton and S. E. Pollack,
``Inside charged black holes I. Baryons'',
Phys.\ Rev.\ D, in press,
gr-qc/0411061
(Paper~1)
(2005).

\bibitem{HKN05}
J. Hansen, A. Khokhlov and I. Novikov,
``Physics of the interior of a spherical, charged black hole with a scalar field'',
%gr-qc/0501015
Phys.\ Rev.\ D\ \textbf{71}, 064013 (2005).

\bibitem{HP96}
S. Hod and T. Piran,
``Critical behavior and universality in gravitational collapses of a charged scalar field'',`
Phys.\ Rev.\ D\ \textbf{55}, 3485--96 (1997).

\bibitem{HP98a}
S. Hod and T. Piran,
``Late-time evolution of charged gravitational collapse and decay of charged scalar hair. I, II, III'',
% development of Price tails outside horizon of charged BH with charged massless scalar field
Phys.\ Rev.\ D\ \textbf{58}, 024017,8,9 (1998).

\bibitem{HP98b}
S. Hod and T. Piran,
``Mass inflation in dynamical gravitational collapse of a charged scalar field'',
Phys.\ Rev.\ Lett.\ \textbf{81}, 1554--7 (1998).

\bibitem{HP99}
S. Hod and T. Piran,
``The inner structure of black holes'',
%gr-qc/9902008
Gen.\ Rel.\ Grav.\ \textbf{30}, 1555--62 (1998).

\bibitem{HO01}
V. Husain and M. Olivier,
``Scalar field collapse in three-dimensional AdS spacetime'',
Class.\ Quant.\ Grav.\ \textbf{18}, L1--9 (2001).

\bibitem{MG03}
J. M. Mart\'{\i}n-Garc\'{\i}a and C. Gundlach,
``Global structure of Choptuik's critical solution in scalar field collapse'',
%gr-qc/0304070
Phys.\ Rev.\ D\ \textbf{68}, 024011 (2003).

\bibitem{MS64}
C. W. Misner and D. H. Sharp,
``Relativistic equations for adiabatic, spherically symmetric gravitational collapse'',
Phys.\ Rev.\ B \textbf{136}, 571--6 (1964).

\bibitem{OP03}
Y. Oren and T. Piran,
``Collapse of charged scalar fields'',
%gr-qc/0306078
Phys.\ Rev.\ D\ \textbf{68}, 044013 (2003).

\bibitem{Ori91}
A. Ori,
``Inner structure of a charged black hole: an exact mass-inflation solution'',
Phys.\ Rev.\ Lett.\ \textbf{67}, 789--92 (1991).

\bibitem{Ori99}
A. Ori,
``Oscillatory null singularity inside realistic spinning black holes'',
%gr-qc/103012
Phys.\ Rev.\ Lett.\ \textbf{83}, 5423--6 (1999).

\bibitem{PI90}
E. Poisson and W. Israel,
``Internal structure of black holes'',
Phys.\ Rev.\ D \textbf{41}, 1796--809 (1990).

\bibitem{Price72}
R. H. Price,
``Nonspherical perturbations of relativistic gravitational collapse. I. Scalar and gravitational perturbations'',
Phys.\ Rev.\ D\ \textbf{5}, 2419--38 (1972).

\bibitem{SP01}
E. Sorkin and T. Piran,
``Effects of pair creation on charged gravitational collapse'',
Phys.\ Rev.\ D\ \textbf{63}, 084006 (2001).

\end{thebibliography}
\end{document}